\documentclass[letterpaper,5p,times,twocolumn]{elsarticle}

 \journal{TexExchange}
 
    \journal{Acta Materialia}
    
    \usepackage{dblfloatfix}
    \usepackage{tgtermes}
    \usepackage{hyperref}
    \usepackage[pagewise]{lineno}
    \modulolinenumbers[1]
    
    \usepackage[section]{placeins}

    \usepackage{hhline}  % http://ctan.org/pkg/hhline
    \usepackage{float}
    \usepackage{subfig}
    \usepackage{multirow}
    \usepackage{graphicx,booktabs}
    \usepackage{xfrac}
    \usepackage{titlesec}
    \usepackage{wrapfig,booktabs}
    \usepackage{mwe}
    \usepackage{adjustbox}
    \usepackage{caption}
    \usepackage{makecell}
    
    \usepackage{amsmath,array}
    \usepackage[table]{xcolor}
    \usepackage{soul}
    \usepackage{array}
    \usepackage{longtable}
    \usepackage{ragged2e}
    \usepackage{mathtools}

    \usepackage{MnSymbol}
    \usepackage{xcolor}
    \usepackage{movie15}

    \usepackage{chngcntr}

    \usepackage{tikz}
    \usetikzlibrary{plotmarks}

    \arrayrulecolor{red}
    \fboxsep=\tabcolsep\relax
    \fboxrule=\arrayrulewidth\relax

    \newcolumntype{A}[2]{%
    >{\minipage{\dimexpr#1\linewidth-2\tabcolsep-#2\arrayrulewidth\relax}\vspace\tabcolsep}%
    c<{\vspace\tabcolsep\endminipage}}
    
    \usepackage{tikz}
    \usepackage{siunitx}

    \graphicspath{{images/}}
     
    \arrayrulecolor{black} % <---

    \biboptions{numbers,sort&compress}

\begin{document}

    %\twocolumn[{
    
    \begin{frontmatter}

    \title{Finite Interface Dissipation Phase Field Modeling of Ni-Nb under Additive Manufacturing Conditions}
    %\title{Rapid Solidification Microstructure of Binary NiNb during Laser Powder Bed Fusion: Non-equilibrium Phase Field Modeling and Experimental Validation}
   
    \author{Kubra Karayagiz$^{a,*,**}$}
    \author{Luke Johnson$^{a}$}
    \author{Raiyan Seede$^{a}$}
    \author{Vahid Attari$^{a}$}
    \author{Bing Zhang$^{c}$}
    \author{Xueqin Huang$^{a}$}
    \author{Supriyo Ghosh$^{a}$}
    \author{Thien Duong$^{b}$}
    \author{Ibrahim Karaman$^{a,b,c}$}
    \author{Alaa Elwany$^{a,c}$}    
    \author{Raymundo Arroyave$^{a,b,c}$}
    \address{$^a$Materials Science \& Engineering Department, Texas A\&M University, College Station, TX 77843}
    \address{$^b$Mechanical Engineering Department, Texas A\&M University, College Station, TX 77843}
    \address{$^c$Industrial \& Systems Engineering Department, Texas A\&M University, College Station, TX 77843}

    \cortext[mycorrespondingauthor]{Corresponding author email: kubra87@tamu.edu} 
    \cortext[mycorrespondingauthor]{Current Affiliation: Mechanical Engineering Department, Worcester Polytechnic Institute, Worcester, MA 01609}    
    \small
    \begin{abstract} 
       %\blindtext[2] 
       During the laser powder bed fusion (L-PBF) process, the built part undergoes multiple rapid heating-cooling cycles, leading to complex microstructures with nonuniform properties. In the present work, a computational framework, which weakly couples a finite element thermal model to a non-equilibrium PF model was developed to investigate the rapid solidification microstructure of a Ni-Nb alloy during L-PBF. The framework is utilized to predict the spatial variation of the morphology and size of cellular segregation structure as well as the microsegregation in single-track melt pool microstructures obtained under different process conditions. A solidification map demonstrating the variation of microstructural features as a function of the temperature gradient and growth rate is presented. A planar to cellular transition is predicted in the majority of keyhole mode melt pools, while a planar interface is predominant in conduction mode melt pools. The predicted morphology and size of the cellular segregation structure agrees well with experimental measurements. 
    \end{abstract}


    \end{frontmatter}
    
    %}]

    %%%%%%%%%%%%%%%%%%%%%%%%%%%%%%%%%%%%%%%%%%%%%%%%%%%%%%%%%%%%%%%%%%%%%
    %%%%%%%%%%%%%%%%%%%%%%%%%%%%%%%%%%%%%%%%%%%%%%%%%%%%%%%%%%%%%%%%%%%%%
    %%%%%%%%%%%%%%%%%%%%%%%%%%%%%%%%%%%%%%%%%%%%%%%%%%%%%%%%%%%%%%%%%%%%%

    \small
    \justifying

    %\linenumbers

    %%%%%%%%%%%%%%%%%%%%%%%%%%%%% main text
    %%%%%%%%%%%%%%%%%%%%%%%%%%%%% main text
    %%%%%%%%%%%%%%%%%%%%%%%%%%%%% main text
    %%%%%%%%%%%%%%%%%%%%%%%%%%%%% main text

    \section{Introduction}\label{sec:intro}
    
    Additive manufacturing (AM) refers to the technologies in which three-dimensional objects are created by adding materials layer-by-layer~\cite{debroy2018additive,frazier2014}. A number of AM processes has been developed over the past few decades. Among them, laser powder bed fusion (L-PBF), a powder-based AM process, has attracted much attention due to its ability to produce fully dense parts with superior mechanical properties. During this process, the material undergoes multiple rapid heating-cooling cycles, leading to complex solidification microstructures with anisotropic properties. Typically, post-processing heat treatments are performed to homogenize and control L-PBF microstructures. However, it is also possible to control the as-deposited microstructure through modulating the manufacturing process parameters, which in turn can reduce the cost and time needed for post-processing and thus facilitate the qualification process \cite{gockel2013understanding}. Therefore, it is essential to develop an understanding of the influence of the process parameters on L-PBF solidification microstructure \cite{debroy2018additive}. Once the solidification microstructure is expressed as a function of process parameters, this knowledge can be used to assist the design of AM materials to obtain desired properties.
    
    Numerical modeling of the solidification microstructure has been performed using various methodologies \cite{yu2018phase,nastac1999numerical,xu2018multiscale,zaeem2013modeling,karma2016atomistic,attari2016phase,attari2018interfacial}. A thorough discussion on solidification microstructures and simulation methods can be found in prior works \cite{doi:10.1080/09506608.2018.1537090,zaeem2015advances,zhu2015modeling,boettinger2000solidification,rappaz2000modelling,stefanescu1995methodologies}. A number of simulation methods has been adopted in the literature to investigate the solidification microstructure of AM parts. Korner et al. \cite{korner2014tailoring} studied the equiaxed and columnar grain structures during electron beam melting (EBM) of IN718 using lattice Boltzmann (LB) method and experimental techniques. It was demonstrated that the grain structure can be tailored from columnar to equiaxed using different scanning strategies. Markl et al. \cite{markl2016numerical} developed a model by coupling LB and cellular automata (CA) methods to investigate the evolution of grain structure during EBM of Ti-6Al-4V. The predicted thermal history in LB model was fed into CA model to predict the grain structure. It was demonstrated that at a low power, stray grains were formed due to incomplete melting.
    
    Nie et al. \cite{nie2014numerical} predicted solidification microstructure during laser AM of a nickel-based superalloy using a multi-scale modeling approach that couples finite element (FE) method and stochastic analysis. The model was used to investigate the evolution of dendritic structure, niobium (Nb) segregation, and formation and morphology of the Laves phase. It was demonstrated that the morphology of the Laves phase, which is an undesired phase, was influenced by the cooling rate. As the cooling rate increased, the morphology of the Laves phase changed from a coarse and chain-like structure to a fine and discrete structure, which is considered to be less detrimental. Lopez-Botello et al. \cite{lopez2017two} employed a CA-FE model to investigate the grain structure (equiaxed vs. columnar) during L-PBF of aluminum alloys. 
    
    Recently, the phase field (PF) method has attracted much attention in investigation of solidification microstructure that results during AM processes \cite{ji2018understanding,francois2017modeling,ghosh2018predictive,wang2019investigation}. This approach is promising with the ability to describe complex microstructural evolution without needing to track the moving interface, as opposed to the classical sharp interface models \cite{boettinger2002phase,ode2001recent,echebarria2004quantitative,karma2001phase,kim2004phase,wang1993thermodynamically,steinbach2013solidification}. The PF model, also called `diffuse interface model', assumes the interface between phases to have a finite thickness, in contrast to the sharp interface assumption. The phase field variable in these models is a state variable with values spanning in space and time, and used to describe the relative amount of a phase. It varies smoothly along the interface with each phase having a constant value. For example, in a system with a solid-liquid interface, the phase field variable takes the value of 0 in the solid phase, 1 in the liquid phase, and a value between 0 and 1 in the interface.

    It has been demonstrated that by weakly coupling PF method with a FE thermal model, quantifiable predictions of the solidification phenomenon during AM can be achieved. Acharya et al. \cite{acharya2017prediction} employed a computational fluid dynamics (CFD) model weakly coupled with a PF model to simulate the microstructure evolution during L-PBF of IN718, which was approximated as a binary alloy. The authors investigated the dendritic structure (size, morphology, and orientation), segregation of Nb, primary dendrite arm spacing (PDAS), and secondary dendrite arm spacing (SDAS). The influence of laser speed on the orientation of dendritic structure was also investigated. 
    
    Keller et al. \cite{keller2017application} investigated the evolution of cellular/dendritic structure and micro-segregation in IN625 alloy during L-PBF process using multiple computational techniques including DICTRA simulation, Scheil-Gulliver model, PF model, and FE-based thermal model. Micro-segregation of multiple substitutional elements (e.g. Nb, Mo, Fe, Cr) were predicted using DICTRA software and Scheil-Gulliver model. Both approaches showed an increase in Nb, Mo, C and a decrease in Fe and Cr at the growth front of $FCC-\gamma$ cells during solidification. The microsegregation of Nb was further investigated using a PF model with an approximation of IN625 alloy to a binary alloy system with the composition of Ni-4 wt.\% Nb. In that work, the predicted PDAS varied between 0.2 $\mu$m and 1.8 $\mu$m depending on the cooling rate, in the range of $10^4$K/s-$10^6$K/s calculated using a FE model. 
    
    Ghosh et al. \cite{ghosh2017primary,ghosh2018simulation} employed a weakly coupled FE-PF model to simulate the solidification microstructure of a Ni-5 wt.\% Nb alloy (a binary approximation of IN718 alloy) during the L-PBF process. Varying values of temperature gradient (G: $2.4\times10^7$ K/m-$0.14\times10^7$ K/m) and growth rate (R: 0.01 m/s - 0.3 m/s) were predicted by the FE model and fed into the PF model to investigate the PDAS and Nb enrichment under different conditions. Predicted PDAS ranged from 0.2 $\mu$m to 0.7 $\mu$m as the cooling rate decreased from $10^6$ K/s to  $3\times10^5$ K/s. Nb segregation in the interdendritic region was predicted as 16 wt.\%. Similar studies in which the solidification microstructure under AM conditions was simulated using a number of different PF models can be found in the literature \cite{geng2017simulation, ghosh2018simulation, kundin2019microstructure,wang2019investigation}. 
    
    Note that the aforementioned PF works were based on the equal diffusion potential condition---i.e., the local equilibrium condition. However, during a typical L-PBF process, the system is out-of-equilibrium due to the extremely high solidification growth rates. Therefore, a PF model with the capability of capturing and describing this rapid solidification phenomenon under L-PBF conditions is required. Recently, Steinbach et al. \cite{steinbach2012phase, zhang2012phase}  proposed a PF model, namely, the finite interface dissipation model, in which both the equilibrium and strongly non-equilibrium conditions can be described successfully. The novelty of this model is that the rate of transport of the components across phase interfaces can be controlled with a kinetic coefficient, namely, ``interface permeability'', $P^{intf}$.  The value of the interface permeability parameter can be chosen (i.e. $P^{intf} \rightarrow 0$) such that the large non-equilibrium case can be modeled. In contrast, when $P^{intf} \rightarrow \infty$, the condition of equal diffusion potential in the conventional models (system with local equilibrium condition) can be recovered. To account for non-equilibrium solidification effects observed during the L-PBF process, the finite interface dissipation PF model is adopted in the present work. Here, it is worth noting that some conventional PF models \cite{kim1999phase} require the solution of additional ancillary equations in order to properly define the structure of the interface, for example, by enforcing the so-called local quasi-equilibrium condition. The finite interface dissipation PF model formulation, on the other hand, eliminates the need of solving these kinds of additional equations, hence the numerical implementation is easier, in principle.

    It is common practice to describe multi-component alloys using a binary approximation due to the complexity of implementing the PF model for multi-component systems. Although it is a reasonable approach to simulate the microstructure of complex alloys consisting of a number of constituent elements (e.g. Inconel 718 with 15 constituent elements), it involves multiple model assumptions that are not necessarily representative of reality, and hence inevitably include inaccuracy and uncertainty in model predictions. For example, NiNb has been widely used as a binary approximation for Ni-based superalloys such as Inconel 718 \cite{ghosh2017primary,acharya2017prediction} and Inconel 625 \cite{keller2017application}. Although the Nb amount in the binary model material is the same as the amount that exists in the approximated multi-component alloy (e.g. 5 wt.\% Nb in Inconel 718), it is evident that there will be contributions from other constituent elements (e.g., Cr, Fe, Mo, Al) in the formation of microstructure phases in the target multi-component alloy, which might influence the accuracy of the comparison with the binary PF model predictions.  
    
    In contrast, the present work provides, for the first time, a consistent framework through using a binary Ni-5 wt.\% Nb (Ni-3.2 at.\% Nb) alloy in both PF simulations and validation experiments. We believe that this will serve to reduce the inaccuracy and uncertainty associated with microstructure simulations when a binary approximation is used in simulations while the complex alloy is used in validation experiments. Indeed, recent modeling and experimental efforts on AM have adopted various binary alloy systems (e.g., Ti-Nb \cite{roehling2018rapid}, Al-Cu \cite{mckeown2016time}, Al-Si \cite{mckeown2016time}, Cu-Ni \cite{Perron_2017}) to elucidate the solidification microstructure developed under AM conditions. 
    
    It is worth noting that the finite interface dissipation PF model adopted in the present work is well suited for describing multi-phase multi-component systems. Very recently, this model was employed in \cite{nomoto1875non} to investigate the solidification microstructure of Fe-Cr-Ni-Mo-C and stainless steels under AM conditions. The authors emphasized the capability of the model in simulating the solidification microstructure for multi-component systems in high temperature gradient and growth rate. 
    
    The present work differs from the above-mentioned study due to the implementation of a FE-PF weakly coupled framework and rigorous experimental validation, as well as the use of the same material for simulations and experimental validation. The main goal of the present work is to present a consistent framework which can be used to elucidate the influence of AM process parameters on the variations in Ni-Nb solidification microstructures during single-track laser melting. Once the microstructure variations in a simple binary alloy during single-track experiments are well understood, the next step is to employ the presented framework to investigate  more complex alloying systems (e.g. Inconel 718) under multi-track multi-layer laser melting conditions.
    
    \begin{figure*}[!ht]
     \centering
     \includegraphics[width=0.9\textwidth]{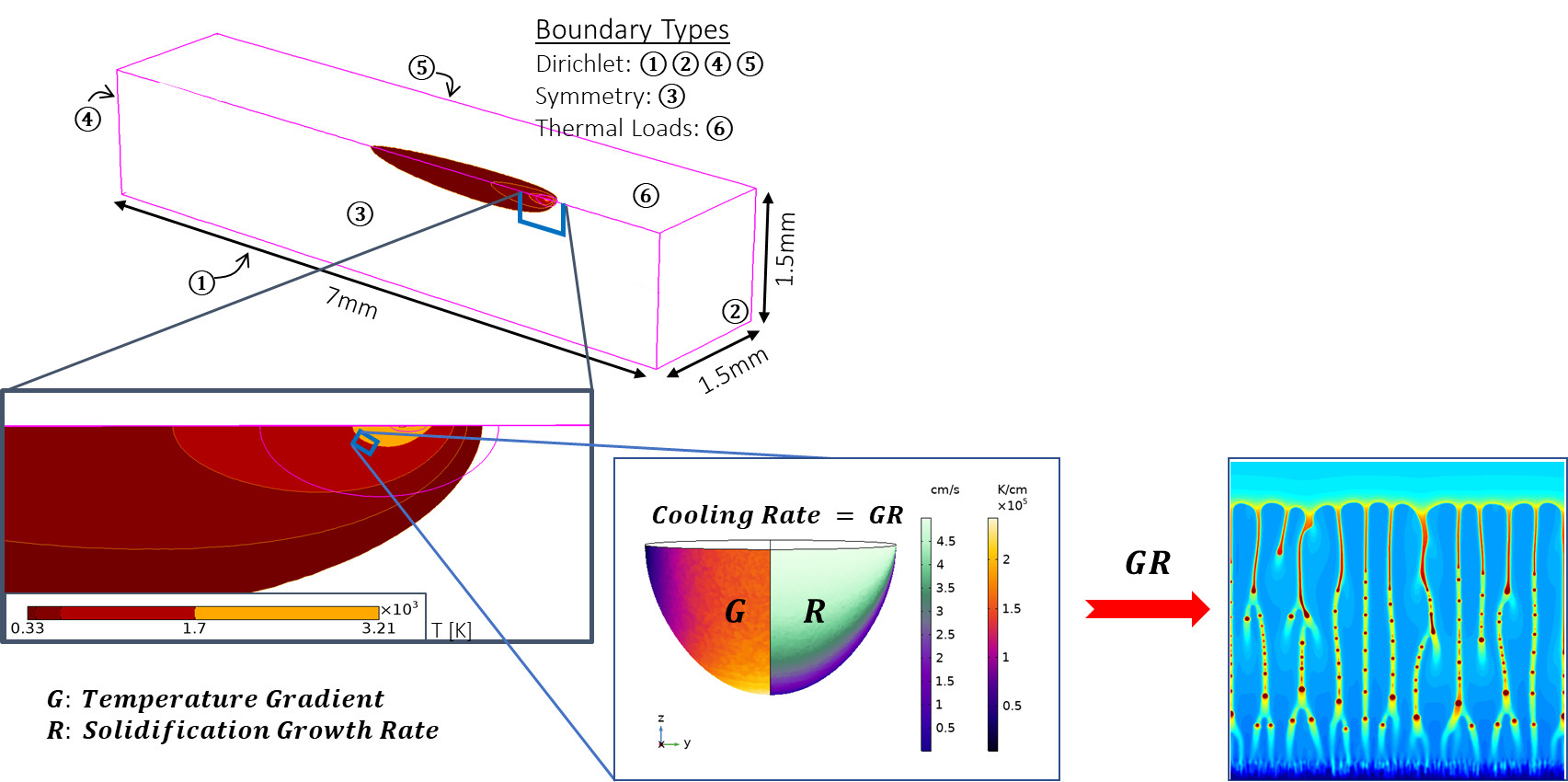}
     \caption{ The weakly coupling procedure of the thermal model with the PF model is represented. An example of predicted heat affected region of Ni-5\%Nb alloy during L-PBF is presented on the three dimensional thermal model geometry. The temperature gradient $G$ and growth rate $R$ are extracted from the thermal model and fed into the PF model to predict the microstructure at the corresponding location at the melt pool indicated by a small rectangle in the magnified view of the heat affected region.}
     \label{fig:fe_pf_coupling}
    \end{figure*}

    Current PF modeling efforts in the AM literature \cite{kundin2019microstructure,ghosh2018simulation,ghosh2018formation,roehling2018rapid} typically report microstructure predictions (e.g. morphology, size of cellular segregation structure, segregation) as a function of temperature gradient ($G$) and growth rate ($R$). Significant insights on the solidification microstructures under AM conditions have been revealed through these types of investigations. However, it is known that the AM microstructures vary locally within a single melt pool even under constant process parameters as well as from melt pool to melt pool depending on the variation of $G$ and $R$ \cite{debroy2018additive}. Therefore, these types of studies fall short of elucidating the relationship between the process parameters and the microstructure variability. In this regard, solidification maps demonstrating AM solidification microstructure as a function of process parameters (e.g. laser power $P$ and scan speed $\vec{v}$) are essential to understand and control the microstructure variations and variability under AM conditions.

    In the present work, we developed an integrated modeling framework which weakly couples an FE-based thermal model to a finite interface dissipation PF model to investigate the rapid solidification microstructure during single track laser melting of a binary Ni-3.2 at.\% Nb alloy. First, the thermal history of the material will be predicted through FE simulations which will then be introduced into the PF model to predict the rapid solidification microstructure (e.g. the morphology, size of cellular segregation structure, and Nb segregation) under L-PBF conditions. To reveal the effect of process parameters on microstructure variability, first, a conventional solidification microstructure map expressing the rapid solidification microstructure as a function of $G$ and $R$ is developed. Next, the microstructure predictions in multiple regions across the melt pool for different combinations of laser power, $P$, and laser scanning speed, $\vec{v}$, are shown and compared with the experimental measurements for validation. %Finally, a solidification microstructure map representing the microstructural variability as a function of $P$ and $\vec{v}$ is demonstrated.\textcolor{blue}{PV map will be added}

    The paper is structured as follows. In Section \ref{sec:exper}, the experimental methodology is described. Section \ref{sec:models} describes the models and procedures. Section \ref{sec:result} presents the experimental measurements, simulation predictions, and discusses the effect of process parameters on the L-PBF microstructures as well as the effect of permeability on the predicted segregation coefficient. A comparison between the results predicted from the finite interface dissipation PF model with those from a conventional PF model \cite{ghosh2017primary} is also provided in Section \ref{sec:result}. Finally, Section \ref{sec:conc} summarizes and addresses the future directions.%need to be re-written!

%%%%%%%%%%%%%%%%%%%%%%%%%%%%% methods
    \section{Experimental Methodology}\label{sec:exper}
    Gas atomized Ni-3.2 at.\% Nb powder was used to manufacture L-PBF NiNb specimens. Single tracks were printed using a 3D Systems ProX DMP 200 Laser Type (fiber laser with a Gaussian profile $\lambda$ = 1070 nm, and beam size = 100 $\mu$m). The tracks were printed on a Ni-3.2 at.\% Nb base plate. These tracks were 10 mm in length with 1 mm spacing between tracks. Cross-sections of the single tracks were wire-cut using wire electrical discharge machining (EDM), and these specimens were polished down to 0.25 $\mu$m with water-based diamond suspension polishing solutions. Kalling's Solution No. 2 (5 g CuCl2, 100 mL HCl, and 100 mL ethanol) was used to etch the Ni-3.2 at.\% Nb single tracks to obtain optical micrographs. 

    Optical microscopy (OM) was carried out using a Keyence VH-X digital microscope equipped with a VH-Z100 wide range zoom lens. Scanning electron microscopy (SEM) and wavelength dispersive spectroscopy (WDS) were performed with a CAMECA SXFive electron probe microanalyzer equipped with a $LaB_6$ electron source. Backscattered electron (BSE) images of polished single tracks were captured at 15 kV and 30 nA. Quantitative WDS composition maps were obtained at settings of 15 kV, 100 nA, and 110 $\mu$m pixel dwell time with a 0.1 $\mu$m step size. Composition data was extracted from the WDS maps to create line scans for visual aid purposes. BSE images were processed using Image J \textregistered software \cite{ws2012imagej} in order to determine PDAS at different locations along select single tracks. The displayed PDAS values were averaged from 30 measurements at each location.

    \section{Model Description}\label{sec:models}

    \subsection{Thermal Model}\label{thermal_model}
    %Numerical simulation of the thermal history during L-PBF was implemented in Comsol Multiphysics \textregistered. 
    The thermal history during the L-PBF process can be determined by numerically solving the transient heat transfer equation given by:
    \begin{equation}\label{eq:heat_transient1}
        \rho C_p \frac{\partial T}{\partial t} + \nabla(-k\nabla T) = Q 
    \end{equation}
    where $\rho$ is the density, $C_p$ is the specific heat, $k$ is the thermal conductivity, $T$ is the temperature, $t$ is the time, and $Q$ is the heat source term. This basic form of the heat transfer equation describes the transient evolution of temperature ($T$) and is typically modified to account for additional physical phenomena that serve to increase the accuracy of the predictions. Examples of such phenomena are phase transformation energy contributions and temperature dependent material properties. In the current model we include phenomena and also perform a coordinate transformation:
    
    \begin{equation}
        \frac{\partial T}{\partial t} = \frac{\partial T}{\partial x}\frac{\partial x}{\partial t} + \frac{\partial T}{\partial y}\frac{\partial y}{\partial t} + \frac{\partial T}{\partial z}\frac{\partial z}{\partial t} = \nabla T \cdot \vec{v} \label{eq:transformation}
    \end{equation}
    
    Substituting Equation \ref{eq:transformation} into Equation \ref{eq:heat_transient1} eliminates the transient portion of the governing heat transfer equation and shifts the reference frame from the material substrate to the heat source which is moving at a constant speed $\vec{v}$.
    
    \begin{equation}
        \rho C_p (\nabla T \cdot \vec{v}) + \nabla(-k\nabla T) = Q \label{eq:heat_steady}
    \end{equation}
    
    The effect of this transformation is the conversion of the transient problem to a steady-state problem, which is a much simpler problem that results in several orders of magnitude decrease in simulation time. This simplification in the governing equations opens up the possibility to use a very fine finite element mesh size (2 $\mu$m) in and around the melt pool, which is essential for the accurate prediction of solidification behavior at the solid-liquid interface. A schematic of the finite element domain and the representative heat affected region can be seen in Fig. \ref{fig:fe_pf_coupling}. 

    Boundary conditions for this model can be broken down into three categories: Dirichlet, Symmetry, and Thermal Loads. Boundaries 1, 2, 4, and 5 are Dirichlet boundary conditions with a fixed temperature of $T_0=298 [K]$. A symmetry condition imposed on the centerline of the track (Boundary 3) essentially halves the computational expense of the simulation. The top surface (Boundary 6) contains all of the heat transfer phenomena that contribute to the source term ($Q$) in Equation \ref{eq:heat_steady}.

    \begin{align}
        Q &= q_{rad} + q_{conv} + q_{vap} + q_{beam} \label{eq:heat_loads}\\
        q_{rad} &= \varepsilon \sigma_B (T_{amb}^4-T^4) \label{eq:radiation}\\
        q_{conv} &= h(T_{amb}-T) \label{eq:convection}\\
        q_{vap} &= L_v \sum_{i=1}^{n} X_i 44.331 p_{i}(T) \sqrt{\frac{MW_{i}}{T}} \label{eq:vaporization}\\
        q_{beam} &= a(T)P\Big[\frac{1}{2\pi\sigma^2}exp\Big(-\frac{(r-r_0)^2}{2\sigma^2}\Big)\Big] \label{eq:beam}
    \end{align}
    
    Equations \ref{eq:radiation}-\ref{eq:beam} describe surface radiation, natural surface convection, vaporization, and deposited beam power, respectively. The radiation, convection, and beam terms are commonly implemented in the finite element modeling of L-PBF. Evaporative energy loss $q_{vap}$ accounts for energy transferred out of the system in the form of mass transfer via hot vapor. The form of this equation is taken from Bolten-Block and Eagar \cite{block1984metal} and modified to include temperature dependent partial pressure relationships for each element. These pressure-temperature relationships are calculated using equations described in \cite{alcock1984vapour}.
    
    In general, the parameters in equations \ref{eq:heat_steady}-\ref{eq:beam} can be sorted into temperature dependent and non-temperature dependent properties. Parameters considered to be constant are: ambient temperature ($T_{amb}$), emissivity ($\varepsilon$), Stefan-Boltzmann constant ($\sigma_B$), molecular weights of Ni and Nb ($MW_i$), laser power ($P$), beam standard deviation ($\sigma$), and beam centerpoint ($r_0$). The group of temperature dependent material properties include: partial pressure ($p_i$), density ($\rho$), specific heat ($C_p$), thermal conductivity ($k$), and laser speed ($\vec{v}$), and absorptivity ($a$).  Phase dependent values for all of these properties can be found in Table \ref{tab:FE_pars}. The natural convection coefficient ($h$) is calculated internally within COMSOL Multiphysics~\cite{comsol} framework based on the domain geometry and orientation.

    \begin{table}[!ht]
    \scriptsize
    \centering
    \caption{Thermophysical and processing parameters used in the FE thermal model. All thermophysical property values were calculated using a weighted average of the pure elemental properties of Ni and Nb.}
    \label{tab:FE_pars}
        \begin{tabular}{lcc}
            % \hline
            \toprule
            \textbf{Phase Properties} & \textbf{Values} & \textbf{Source} \\ \toprule
            % \hline
            % \hline
            \textbf{Solid}  &  &  \\
            $\rho_S$  & 8900 [kg/m$^{3}$] & MSDS \\
            $k_S$  & 85 [W/mK] & \cite{gale2003smithells}  \\
            $Cp_S$  & 550 [J/kgK] & \cite{gale2003smithells}  \\
            $a_S$  & 0.3 [unitless] & \cite{trapp_situ_2017} \\ \midrule
            % \hline
            \textbf{Solid} $\leftrightarrow$ \textbf{Liquid} & & \\ 
            $T_m$ & 1703 [K] & MSDS \\
            $\Delta T_m$ & 50 [K] & \cite{gale2003smithells} \\
            $L_m$ & 2.9(10)$^{5}$ [J/kg] & \cite{chase1998nist} \\ \midrule
            % \hline
            \textbf{Liquid}  &   &  \\ 
            $\rho_L$  & 8450 [kg/m$^{3}$] & \cite{gale2003smithells} \\
            $k_L$  & 120 [W/mK] & \cite{gale2003smithells}  \\
            $Cp_L$  & 650 [J/kgK] & \cite{gale2003smithells}  \\
            $a_L$  & 0.3 [unitless] & \cite{trapp_situ_2017} \\ \midrule
            % \hline
            \textbf{Liquid} $\leftrightarrow$ \textbf{Vapor} & & \\
            $T_v$ & 3209 [K] & \cite{chase1998nist} \\
            $\Delta T_v$ & 200 [K] & Approx. \\
            $L_V$ & 7.1(10)$^{6}$ [J/kg] & \cite{chase1998nist} \\ \midrule
            % \hline
            \textbf{Vapor}  &  &  \\
            $\rho_V$  & Temp. Dep. & \cite{block1984metal} \\
            $k_V$  & 1000 [W/mK] & Approx. \\
            $Cp_V$  & Temp. Dep. & \cite{chase1998nist} \\
            $a_V$  & 0.6 [unitless] & \cite{trapp_situ_2017} \\
            $p_i$ & Temp. Dep. & \cite{alcock1984vapour} \\ \toprule
            % \hline
            % \hline
            \textbf{Constants} & \textbf{Values} & \textbf{Source} \\ \toprule
            % \hline
            % \hline
            \textbf{Laser} & & \\
            $P$ & 70-255 [W] & User \\
            $\vec{v}$ & 50-2300 [mm/s] & User \\
            $4\sigma$ & 70 [$\mu$m] & Manuf. \\ \midrule
            % \hline
            \textbf{General}  & & \\
            $T_{amb}$ & 298 [K] & \\
            MW$_{Ni}$ & 58.7 [g/mol] & \\
            MW$_{Nb}$ & 92.9 [g/mol] & \\
            $\varepsilon$ & 0.7 & \\
            % \hline
        \end{tabular}
    \end{table}
    
    The phase-dependent property values in Table \ref{tab:FE_pars} are calculated using an average of the elemental properties for each constituent \cite{gale2003smithells}, weighted by the corresponding atomic percent. This rule-of-mixtures approximation is necessitated by the lack of experimental thermophysical property measurements for this custom NiNb alloy. A weighted average is sufficient in this case because the alloy is a very dilute single-phase solid-solution. In particular, the absorptivity values were chosen to approximate a recent study by Trapp et al. \cite{trapp_situ_2017} which shows experimental evidence for low effective absorptivity of the solid/liquid phases and high effective absorptivity upon vaporization and keyhole formation. Additionally, a powder layer with effective material properties is not directly modeled due to the fact that its primary effect (increased laser absorptivity) is negligible at steady-state conditions when the laser is solely incident upon the solid, liquid and vapor phases.
    
    Smooth transitions between the phase-dependent thermophysical property values in Table \ref{tab:FE_pars} are accomplished by averaging the properties of each phase based on their respective fractions during the transformation. Latent heat contributions for melting and vaporization ($L_m$ and $L_v$) are included in the model by adding the appropriate term directly to the heat capacity value during their respective transformations. Details of this effective property approach can be found in \cite{karayagiz2019numerical}.

    \subsection{Phase Field Model with Finite Interface Dissipation}\label{PFM_sec}

    The phase field model with finite interface dissipation introduced by Steinbach et al. \cite{steinbach2012phase,zhang2012phase} is adopted in the present work to investigate the rapid solidification process during L-PBF of Ni-3.2 at.\% Nb alloy. The model has been proven to be capable of modeling non-equilibrium solidification behavior. Here, a brief description of the model is presented. Further details on full derivation can be found in the references provided above.
    
    The derivation of a typical PF model starts with the description of the free energy functional that may consist of local chemical free energy density, interfacial energy density, elastic energy density, electrochemical energy density depending on the studied problem \cite{moelans2008introduction,hu2001phase,yurkiv2018phase,yi2018strain,attari2019exploration}. In the present work, the free energy functional has contributions from the local chemical free energy density ($f^{chem}$) and from the interfacial energy density ($f^{intf}$):
    
    \begin{equation}
        F^{tot} = \int_{\Omega} \left ( f^{intf} + f^{chem} \right) d\Omega
    \end{equation}
    
    \begin{equation}
        f^{intf} = \frac{4\sigma_{\alpha\beta}}{\eta} \left \{ -\frac{\eta^2}{\pi^2} \nabla \phi_\alpha \cdot \nabla\phi_\beta + \phi_\alpha \phi_\beta \right \}
    \end{equation}
    
    \begin{equation}
        f^{chem} = \phi_{\alpha} f_{\alpha} ( c_\alpha ) + \phi_{\beta} f_{\beta} ( c_\beta ) + \lambda \{ c - \left ( \phi_{\alpha} c_{\alpha} + \phi_{\beta} c_{\beta} \right) \}
    \end{equation}

    \noindent
    where $\sigma_{\alpha\beta}$, $\eta$, $\phi_{\alpha/\beta}$, $c_{\alpha/\beta}$, and $c$ are the interfacial energy, the interface width, the phase fractions of $\alpha/\beta$ phases, the phase concentrations of $\alpha/\beta$ phases, and the overall concentration, respectively. The summation of the phase fractions is 1 throughout the system with the relationship of $\phi_{\alpha}+\phi_{\beta}=1$. $\lambda$ is the Lagrange multiplier, which is introduced to assure the solute conservation constraint given by: $c=c_{\alpha}\phi_{\alpha}+c_{\beta}\phi_{\beta}$. $f_{\alpha}$ and $f_{\beta}$ are the free energy densities of the corresponding phases and expressed within the CALPHAD formalism \cite{saunders1998calphad}. At the selected alloy composition of $c_{alloy}=3.2\:(at.\% Nb)$, two phases (liquid and $\gamma$) can be described as:
    %$c_{alloy}=0.032 (at.\:frac.\:Nb)$
   % \footnotesize

    \begin{multline}\label{eqn:gibbs}
     f_{\alpha}V_m = c_{\alpha}G^{0}_{Nb} + 
     (1-c_{\alpha})G^{0}_{Ni} \\+ RT \left( c_{\alpha}ln(c_{\alpha})+ (1-c_{\alpha})ln(1-c_{\alpha}) \right)\\+c_{\alpha}(1-c_{\alpha})\sum_{i=0}^{n} G_i (2c_{\alpha} - 1)^i  
    \end{multline}

    \noindent    
    where $V_m$, $R$ and $T$ represent the molar volume, ideal gas constant and temperature, respectively. $G_{Nb}^0$, $G_{Ni}^0$ are reference states of constituent elements. $G_i$ terms are coefficients contributing to excess Gibbs energy. The coefficients for each phase (i.e. $G_{Nb}^0$, $G_{Ni}^0$, and $G_i$) are obtained from \cite{joubert2004assessment}.
    
    The evolution equations of the phase concentrations, $c_{\alpha}$ and $c_{\beta}$, are derived through variational principles. The final form of the evolution equations of the phase concentrations:
    
    \begin{equation}
        \phi_{\alpha} \dot{c}_{\alpha} = \nabla(\phi_\alpha D_\alpha \nabla c_\alpha) + P^{intf} \phi_\alpha \phi_\beta \left( \frac{\partial f_\beta}{\partial c_\beta} - \frac{\partial f_\alpha}{\partial c_\alpha}  \right) + \phi_\alpha \dot{\phi}_{\alpha} \left( c_\beta - c_\alpha \right)
        \label{eq:concentration_alp}
    \end{equation}
    
    \begin{equation}
        \phi_{\beta}\dot{c}_{\beta} = \nabla(\phi_\beta D_\beta \nabla c_\beta) + P^{intf} \phi_\alpha \phi_\beta \left( \frac{\partial f_\alpha}{\partial c_\alpha} - \frac{\partial f_\beta}{\partial c_\beta} \right) + \phi_\beta \dot{\phi}_{\beta} \left( c_\alpha - c_\beta \right)
        \label{eq:concentration_bet}
    \end{equation}
    
    \noindent
    where $D_{\alpha}$, $D_{\beta}$ are the chemical diffusivities in the $\alpha$ and $\beta$ phases, respectively, and $P^{intf}$ is the interface permeability defined as: $P^{intf}=\frac{8M}{a\eta}$. Here, $M$ is the atomic mobility and $a$ is the lattice constant. Further information on the physical meaning of the interface permeability, $P^{intf}$, can be found in the referenced papers \cite{steinbach2012phase,zhang2012phase}. 
    
    Similarly, the evolution equations of the phase fractions, $\phi_{\alpha}$ and $\phi_{\beta}$, are derived through variational principles. The final evolution equation for the phase fraction $\phi_{\alpha}$ is:
    
    \begin{equation}\label{eqn:pfm_eqn}
        \dot{\phi_{\alpha}} = K \left \{ \sigma_{\alpha\beta} [ \nabla^2 \phi_\alpha + \frac{\pi^2}{\eta^2} (\phi_\alpha - \frac{1}{2} ) ] - \frac{\pi^2}{8\eta} \Delta g^{\phi}_{\alpha\beta} \right \}
    \end{equation}
    
    \begin{equation}\label{eq:permeability}
        K = \frac{8P^{intf}\eta\mu_{\alpha\beta}}{8P^{intf}\eta + \mu_{\alpha\beta} \pi^2 (c_\alpha - c_\beta)^2 }
    \end{equation}
    
    \begin{equation}
        \Delta g_{\alpha\beta}^{\phi} = f_\alpha - f_\beta + \left (\phi_\alpha \frac{\partial f_\alpha}{c_\alpha} - \phi_\beta \frac{\partial f_\beta}{c_\beta} \right) (c_\beta - c_\alpha)
    \end{equation}
    
    \noindent
    where, $\mu_{\alpha\beta}$ is the interfacial mobility, $K$ is the kinetic coefficient describing the effect of finite diffusion and redistribution at the interface and $\Delta_{\alpha\beta}^{\phi}$ is the chemical driving force. The evolution equation for $\phi_{\beta}$ can be easily obtained using the relationship: $\phi_{\beta}=1-\phi_{\alpha}$. Therefore, it is not explicitly demonstrated here.
    
    In order to model the cellular/dendritic structure, the phase field equation (Eq. \ref{eqn:pfm_eqn}) should be modified properly. Following the simplification suggested in \cite{steinbach2009phase}, the interfacial energy $\sigma_{\alpha\beta}$ and interface mobility $\mu_{\alpha\beta}$ terms are modified to their anisotropic forms as $\sigma_{\alpha\beta}^*(\vec{n})$ and $\mu_{\alpha\beta}^*(\vec{n})$, where $\vec{n}$ is the interface normal vector described as $\vec{n}=|\nabla\phi_\alpha|/\nabla\phi_\alpha$, in which $\alpha$ represents the solid phase. More explicitly, these two terms are described as: $\sigma^{*}_{\alpha\beta} (\vec{n}) = \sigma_{\alpha\beta}(\vec{n}) + \sigma^{''}_{\alpha\beta} (\vec{n})$,  
    $\mu^{*}_{\alpha\beta} (\vec{n}) = \mu_{\alpha\beta} (\vec{n}) + \mu^{''}_{\alpha\beta} (\vec{n})$, where $\sigma^{''}_{\alpha\beta} (\vec{n})$ and $\mu^{''}_{\alpha\beta} (\vec{n})$ are the second derivatives of  $\sigma^{''}_{\alpha\beta} (\vec{n})$ and $\mu^{''}_{\alpha\beta} (\vec{n})$ with respect to ($\vec{n}$). For the solid-liquid interface with 4-fold anisotropy, $\sigma_{\alpha\beta} (\vec{n})$ and $\mu_{\alpha\beta} (\vec{n})$ are approximated as:

    \begin{equation}
        \sigma_{\alpha\beta} (\vec{n}) = \sigma^{0}_{\alpha\beta} \left (1 - \epsilon [3-4(n_x^4+n_y^4]) ] \right )
    \end{equation}
    
    \begin{equation}
        \mu_{\alpha\beta} (\vec{n}) = \mu^{0}_{\alpha\beta} \left (1 - \epsilon [3-4(n_x^4+n_y^4]) ] \right )
    \end{equation}
    
    \noindent
    where $\sigma^{0}_{\alpha\beta}$, $\mu^{0}_{\alpha\beta}$ and $\epsilon$ are the interfacial energy coefficient, interface mobility coefficient and anisotropy coefficient, respectively. $n_x$ and $n_y$ represent the $x$ and $y$ components of the norm $\vec{n}$.
 
    To account for the varying temperature during solidification process, the frozen temperature approach is employed. It neglects the latent heat release during solidification and assumes a constant temperature gradient G. The temperature field along $y$ axis is calculated as:
    
    \begin{equation}
        T(y) = T_0 + G(y-Rt)
    \end{equation}

    \noindent    
    where $T_0$ is the reference temperature, $R$ is the growth rate and $t$ is the time. This approach has been widely adopted in the literature to model directional solidification \cite{echebarria2004quantitative, takaki2014two}.
    
    \subsection{Computational Procedures}\label{Computation} 
    \subsubsection{Macroscopic Thermal Model}
    Numerical simulation of the thermal history during L-PBF was implemented in COMSOL Multiphysics \textregistered~\cite{comsol}. A fine element with the size of (2 $\mu$m) was adopted in and around the melt pool, while relatively coarser elements were utilized at the further locations. The domain size was set to $7mm \times 1.5mm \times 1.5mm$ which is sufficiently large enough to negate boundary effects for any combination of power and speed simulated in this work.
        
    Single track laser melting simulations were run at varying laser power ($P$: 70-255 [W]) and laser speed ($\vec{v}$:  50-2300 [mm/s]). The values of $G$ and $R$ parameters were extracted from the these simulations to use as inputs for the PF model as illustrated in Fig. \ref{fig:fe_pf_coupling}. Both $G$ and $R$ are calculated on the trailing half of the melt pool which is the portion subject to solidification as the melt pool travels through the substrate at a constant velocity. As suggested in the frozen temperature approach, G is assumed to be constant in the simulation domain of the PF model and calculated from the partial derivatives of temperature with respect to each Cartesian coordinate with:
    
    \begin{equation}
        G=\sqrt{\Big(\frac{\partial T}{\partial x}\Big)^2+\Big(\frac{\partial T}{\partial y}\Big)^2 + \Big(\frac{\partial T}{\partial z}\Big)^2}
    \end{equation}
    
    The growth rate is geometrically derived as the projection of laser velocity $\vec{v}$ onto the normal vector of the solidification front using the angle ($\theta$) between said vectors \cite{wei_evolution_2015,lienert_fundamentals_2011}:
    
    \begin{equation} \label{eq:growth_rate}
        R = |\vec{v}| \cdot \cos{\theta}
    \end{equation}

    \begin{table}[!ht]
        \scriptsize
        \centering
        \caption{Material parameters used in the PF simulations}
        \label{tab:mat_pars}
        \begin{tabular}{lr} \toprule
            Parameters                         & Values \\ \midrule
            Grid spacing, $\Delta x$ ($nm$) & 8 \cite{ghosh2017primary} \\
            Interface width, $\eta$ ($nm$)     & $32$ \cite{ghosh2017primary}\\
            Molar volume, $V_m$ ($cm^3/mol$)     & 6.59 \cite{campbell2009high}\\
            Interface energy,  $\sigma_{\alpha\beta}$ ($J/cm^2$) & $1.3\times10^{-5}$ \\
            Interface mobility,  {$\mu_{\alpha\beta}$} ($cm^3/Js$) & 1 \\
            Diffusivity of solid,  $D_S$ ($cm^2/s$) & $1.0\times10^{-8}$ \cite{ghosh2017primary} \\
            Diffusivity of liquid,  $D_L$ ($cm^2/s$) & $3.0\times10^{-5}$  \cite{ghosh2017primary} \\ 
            Interface permeability,   $P^{intf}$ ($cm^3/Js$) & 8333 \\ 
            Lattice constant,  $a$ ($cm$) & $3.0\times10^{-8}$ \\ 
            Atomic mobility, $M$ ($cm^5/Js$) & $1.0\times10^{-10}$ \\ 
            Anisotropy coefficient, $\epsilon$     & 0.03 \cite{ghosh2017primary}\\ 
            Equilibrium freezing range, $\Delta T$ ($K$) & 14 \cite{joubert2004assessment} \\
            Equilibrium segregation coefficient, $ke$ &  0.68 \cite{joubert2004assessment} \\\bottomrule
        \end{tabular}
    \end{table}

    \subsubsection{Finite Interface Dissipation Phase Field Model}
    The phase-field and concentration evolution equations, \ref{eq:concentration_alp}, \ref{eq:concentration_bet}, \ref{eqn:pfm_eqn} are numerically solved using the finite difference method (forward in time, centered in space). A dynamic time step was adopted in the present work to ensure the numerical stability. The implemented dynamic time step algorithm enforces that the change in the concentration ($c_{\alpha/\beta}$) or 
    phase field variables ($\phi_{\alpha/\beta}$) do not exceed 0.001, which is a conservative value.  Here, the change is given by $\dot{z}\times \Delta t$ where $\dot{z}$ is the rate of change in variable $z$ ($z$: $c_{\alpha/\beta}$ or $\phi_{\alpha/\beta}$), and $\Delta t$ is the time step. $\Delta t$ is adjusted at each iteration depending on the value of $\dot{z}$. To prevent instability, a small $\Delta t$ is enforced for large $\dot{z}$, indicating a high driving force. On the other hand, a relatively larger $\Delta t$ is allowed for small $\dot{z}$, which enables faster computation as the driving force decreases. The numerical experiments revealed the maximum time step that ensures numerical stability being approximately $\Delta t=1.4\times10^{-11} s$, if a constant time step is adopted in the simulations. It was found out that, the overall performance of the implemented dynamic time step algorithm is much better since it increases the time step up to an order of magnitude higher than that of the value suggested from numerical experiments, as the driving force to evolve the system decreases. While this helped to accelerate the simulations to some extent, primarily it was achieved by adopting OpenMP parallelization approach.
    
    Here, it is worth noting that the phase field model is ill-defined for $\phi_{\alpha/\beta}=0$ and $\phi_{\alpha/\beta}=1$. To handle this, we use a threshold value in the code and check whether the calculated value of $\phi_{\alpha/\beta}$ being above or below the allowed range which is $[0.00001,0.9999]$. If the calculated value is beyond the threshold, it is then reset to the allowed maximum/minimum.
    
    To investigate the general features of the microstructure, a 2-dimensional simulation domain with the size of $616\Delta x$ by $4500\Delta y$ is utilized, where the grid spacing $\Delta x=\Delta y=0.008 \mu m$, resulting in a physical size of $\sim5 \mu m$ by $36 \mu m$, while a relatively smaller simulation domain with the size of $800\Delta x$ by $800\Delta y$ is adopted to study the effect of process parameters. No-flux boundary conditions are applied to all boundaries. The initial simulation domain consists of a thin layer of FCC-$\gamma$ solid at the bottom and a thick layer of liquid on the top. Initially, random perturbations are applied to the solid-liquid interface to promote cellular segregation structure. Initial Nb compositions of the solid and liquid are set to $c_s^0=c_s^{eq}=2.2\:(at.\% Nb)$ at ($T=T_0=1695K $) and $c_l^0=c_{alloy}=3.2\:(at.\% Nb)$. 
     
     The material parameters used in the simulations are listed in Table \ref{tab:mat_pars}. Note that in the present work, the parameters employed in Ghosh et al. \cite{ghosh2017primary} have been adopted as much as possible to be able to compare the results from the finite interface dissipation PF model with those predicted by the conventional PF model as presented in \cite{ghosh2017primary}. Here, it is worth noting that the value of M in the present work is estimated according to $D_{\alpha}=\frac{M R T}{Vm}$ \cite{duong2015integrated}, where $D_\alpha$ refers to $D_S$ or $D_L$. Once the values of the parameters are plugged into this equation, the value of $M$ is found to be within the range of $10^{-12}-10^{-9} [cm^5/Js]$, where the lower and upper limits are determined by $D_S$ and $D_L$, respectively. Considering the given limits, the value of $M$ is set to $1.0 \times 10^{-10} cm^5/Js$ in the present work.

     The chemical free energies of each phase are expressed within the CALPHAD formalism \cite{saunders1998calphad} as described in Eq. \ref{eqn:gibbs}, using the thermodynamic data given in \cite{joubert2004assessment}. The phase diagram of the binary Ni-Nb alloy created in Thermocalc Software utilizing the data from \cite{joubert2004assessment} can be seen in Fig. \ref{fig:phase_diagram}.

     The frozen temperature approach as described above is used to model the influence of the temperature. Varying values of temperature gradient, $G$, and growth rate, $R$, obtained from the thermal model, with the order of magnitudes varying from $10^6 K/m$ to $10^8 K/m$, and $10^{-3}  m/s$ to $1 m/s$, respectively, are fed into the PF model to investigate the morphology, size of cellular segregation structure, and segregation of the simulated microstructures.
   
    \section{Results and Discussion}\label{sec:result}

     Energy density is often employed as a metric to determine the printability range of L-PBF manufactured components. A number of energy density formulations using different process parameters (e.g., laser power, laser speed, hatch spacing, layer thickness, laser beam diameter) have been defined and adopted in the current literature. A thorough discussion on the success and limitation of the energy density formulations as a design parameter can be found elsewhere \cite{bertoli2017limitations,mahmoudi2018printability}. Keeping those limitations in mind, for simplification we use the linear energy density (LED)-based categorization (e.g. low LED, high LED) while discussing the effect of laser power ($P$) and speed ($\vec{v}$) on the melt pool characteristics (e.g. melt pool size, geometry, temperature gradient, cooling rate) and microstructure (e.g. morphology, size of cellular segregation structure, segregation). LED in this work is calculated by:
    
        \begin{equation}
         LED=\frac{P}{\vec{v}}
        \end{equation}    
        
    \subsection{Macrostructure Characterization: Experiments} 
    
\begin{figure*}[!hbt]
     \centering
     \includegraphics[width=1\textwidth]{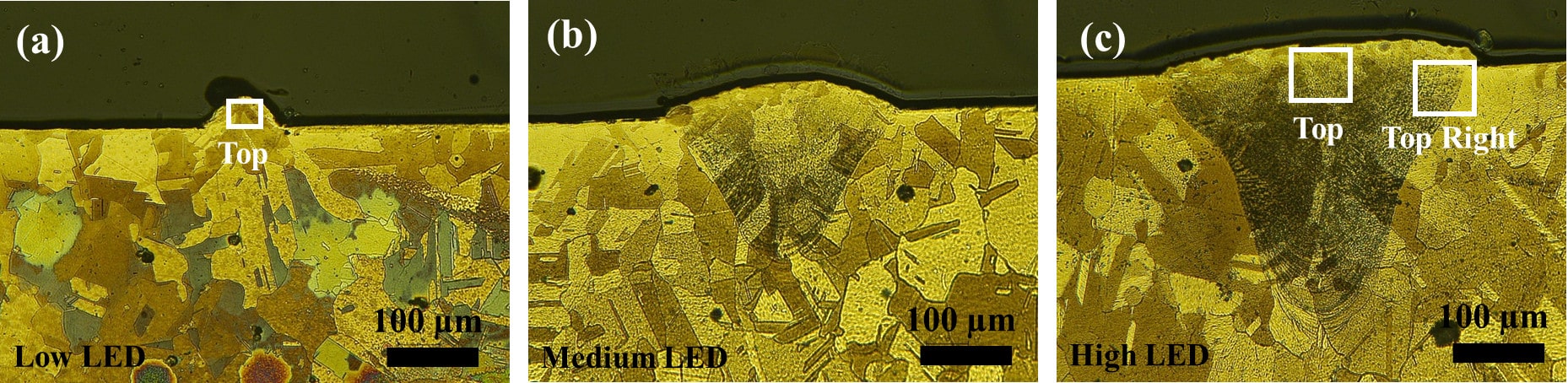}     
     \caption{Optical micrographs demonstrating transverse cross-sections of melt pool obtained under (a) $P$: 162 W,  $\vec{v}$: 957 mm/s, LED (low): 0.169 J/mm (b) $P$: 96 W,  $\vec{v}$: 67 mm/s, LED (medium): 1.43 J/mm (c) $P$: 122 W,  $\vec{v}$: 50 mm/s, LED (high): 2.44 J/mm. From left, to right a clear transition from conduction mode to keyhole mode is shown.}
     \label{fig:experiment_melt_pool}
    \end{figure*}

    For the experimental validation, three sets of process parameters are selected. Figure \ref{fig:experiment_melt_pool} shows optical micrographs of transverse cross-sections of single track laser melts using these three parameter sets: (a) $P$: 162 W,  $\vec{v}$: 957 mm/s, LED (low): 0.169 J/mm (b) $P$: 96 W,  $\vec{v}$: 67 mm/s, LED (medium): 1.43 J/mm (c) $P$: 122 W,  $\vec{v}$: 50 mm/s, LED (high): 2.44 J/mm as listed in Table \ref{tab:proc_param}. The optical micrographs demonstrate a variation in the melt pool size and geometry depending on variation in the values of LED. The melt pool widths are measured as $98\pm1 \mu m$ , $288\pm9 \mu m$  , and $354\pm11 \mu m$, while the melt pool depths are measured as $49\pm2 \mu m$, $178\pm6 \mu m$, and $322\pm4 \mu m$, respectively for low, medium and high LED cases as listed in Table \ref{tab:melt_dim}. From left to right in Fig. \ref{fig:experiment_melt_pool}, an increase in both melt pool width and depth is revealed due to the increase in the LED. Note that the experimental uncertainty denotes one standard deviation from the mean.
    
    Typically there are three laser heating modes that influence the melt pool geometry: conduction mode, transition keyhole mode, and keyhole (penetration) mode. Conduction mode heating occurs at low LED and is characterized by wide and shallow melt pool shape. Transition keyhole mode occurs at medium LED and results in a melt pool with an aspect ratio (depth/width) of around 1. Keyhole mode heating occurs at high LED and is characterized by deep melt pools with a large aspect ratio typically greater than 1.5 \cite{roehling2018rapid}. The optical micrographs reveal a clear trend from conduction to transition keyhole mode heating as the LED increases from left to right in Fig. \ref{fig:experiment_melt_pool}. Note that, within the range of $P: 70-255$ [W] and $\vec{v}: 50-2300$ [mm/s] used in the present work, conduction mode heating was achieved in the majority of single track experiments. All the successful prints with $P<100$ W or $\vec{v}>957$ mm/s revealed conduction mode heating, while varying heating modes were achieved depending on the combination of $P$ and $\vec{v}$. It is apparent from these results that a variation in process parameters leads to a variation in melt pool characteristics, and in turn the solidification microstructure.    
    
        \begin{table}[!ht]
        \scriptsize
        \centering
        \caption{Process Parameters used in the Single-Track Laser Melting Experiments}
        \label{tab:proc_param}
        \begin{tabular}{ccc} \toprule
            Laser Power [W] & 
            Laser Speed [mm/s] &
            Linear Energy Density [J/mm]\\ \midrule
            162 & 957 & 0.169 (low)\\
            96 & 67 & 1.43 (medium) \\
            122 & 50 & 2.44 (high)\\\bottomrule
        \end{tabular}
    \end{table}
    
    \begin{table}[!ht]
        \scriptsize
        \centering
        \caption{Measurements of the Melt Pool Dimensions under Varying Processing Conditions as listed in Table \ref{tab:proc_param}}
        \label{tab:melt_dim}
        \begin{tabular}{ccc} \toprule
            Melt Pool Width [$\mu m$] &
            Melt Pool Depth [$\mu m$] &
            Linear Energy Density [J/mm]\\ \midrule
            $98\pm1$ & $49\pm2$ & 0.169 (low)\\
            $288\pm9$ & $178\pm6$ &  1.43 (medium) \\
            $354\pm11$ & $322\pm4$  & 2.44 (high)\\\bottomrule
            
        \end{tabular}
    \end{table}

    % UPDATED FIGURE %
    \begin{figure*}[!ht]
     \centering
     \includegraphics[width=1.05\textwidth]{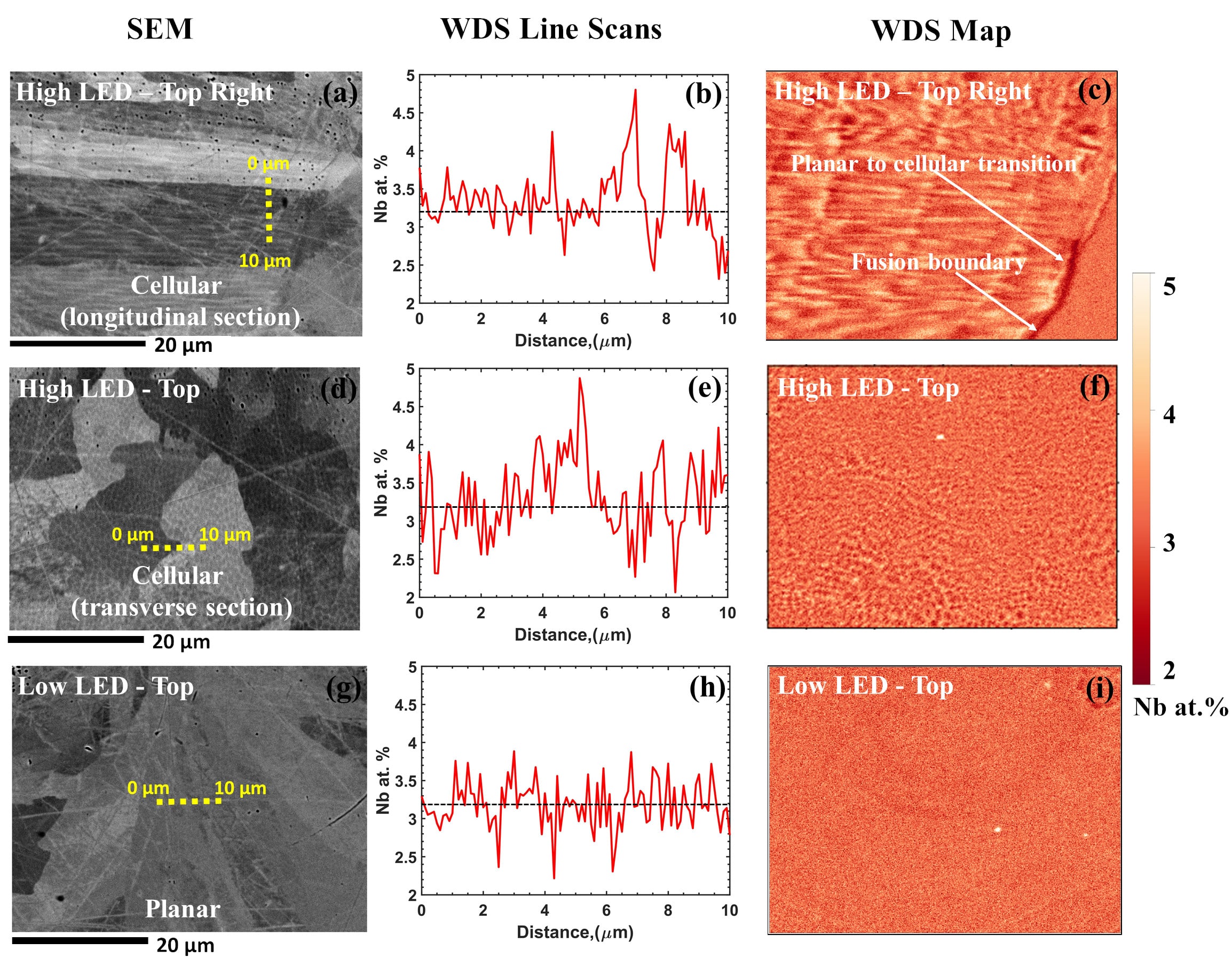} 
     \caption{The microstructure of the selected rectangular regions in Fig. \ref{fig:experiment_melt_pool} are characterized using SEM (left column), WDS line scans(middle  column), and WDS maps (right  column). SEM images show cellular and planar morphologies for the corresponding LED conditions in Fig. \ref{fig:experiment_melt_pool}.}
     \label{fig:experiment_wds_sem}
    \end{figure*} 
    
    %\hl{~(UPDATED FIGURE)}

    \subsection{Microstructure Characterization: Experiments} 
    Multiple regions from the melt pools indicated with small rectangles in Fig. \ref{fig:experiment_melt_pool} are selected and characterized using SEM and WDS techniques as shown in Fig. \ref{fig:experiment_wds_sem}. SEM micrographs of the selected regions reveal two types of growth structures: cellular and planar. 
    
    The SEM micrograph (Fig.~\ref{fig:experiment_wds_sem} (a)) corresponding to the ``Top Right'' position under ``High LED'' (see Fig. \ref{fig:experiment_melt_pool} (c)) presents a cellular structure growing normal to the fusion boundary and antiparallel to the heat flux direction. The appearance of columnar cells indicates the directional growth in a positive temperature gradient \cite{kurz1989fundamentals}. Cells are also visible in the WDS maps of the corresponding region (as shown in Fig. \ref{fig:experiment_wds_sem} (c)) due to composition differences between the cell cores and walls. This variation in the composition is the result of solute rejection by the growing cells. The WDS map indicates that cell walls are Nb-rich while cell cores are Ni-rich. Cell growth rates can vary from below the limit of constitutional supercooling to beyond the limit of absolute stability \cite{kurz1989fundamentals}.

    It is worth noting that the fusion boundary with Nb depletion in the WDS map (see Fig.~\ref{fig:experiment_wds_sem} (c)) is indicative of the local equilibrium planar structure. The appearance of planar structure near the fusion boundary can be explained due to the presence of extremely low growth rates. Solidification starts near the fusion boundary once the temperature is below the liquidus temperature (1701 K for the alloy composition of 3.2\:(at.\% Nb) used in the present work \cite{joubert2004assessment}). According to the phase diagram \cite{joubert2004assessment}, the equilibrium composition of the solid phase just below the solidus temperature is around 2.2\:(at.\% Nb), which agrees well with the low Nb composition shown in the planar region at the WDS map. This planar structure quickly transitions to a cellular structure as the temperature decreases below the liquidus temperature and the growth rate exceeds the limit of constitutional supercooling. The constitutional supercooling criteria will further be discussed in the following subsections.
    
\begin{figure}[!ht]
     \centering
     \includegraphics[width=1\columnwidth]{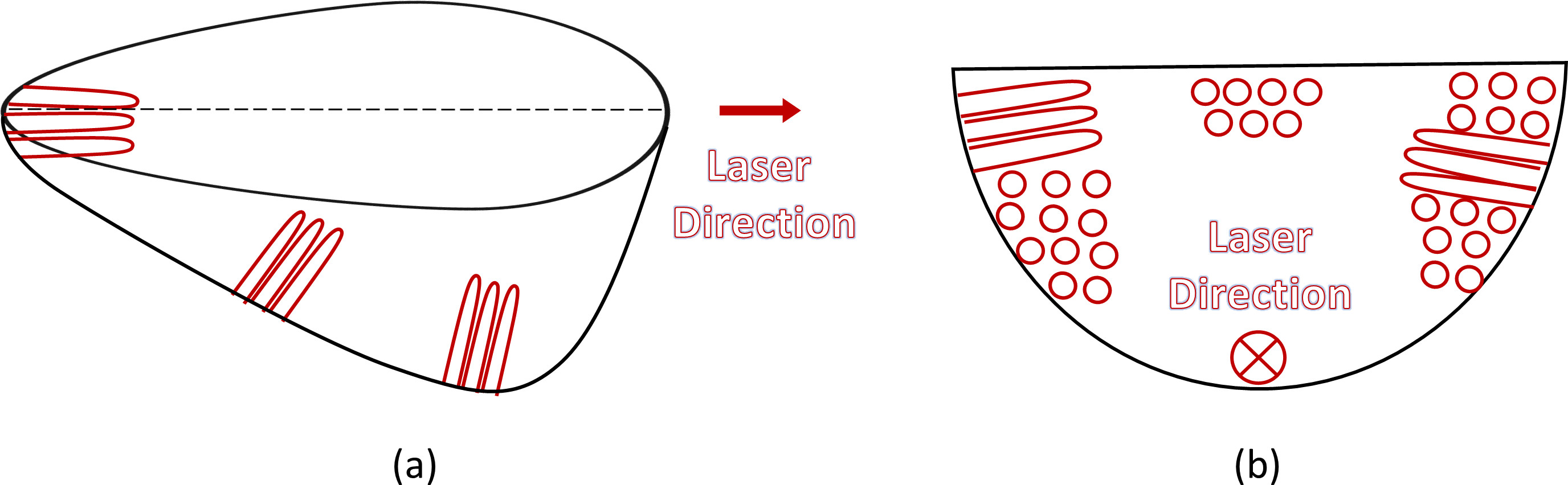}
     \caption{The formation of cellular structures is shown in (a) longitudinal section (b) transverse section of melt pool. Depending on the orientation of the cells, two different view of cellular structure (columnar and circular) is observed in the transverse section of melt pool.}
     \label{fig:cell_illustration}
    \end{figure}

    Cellular segregation structure is also revealed in the SEM micrograph (Fig. \ref{fig:experiment_wds_sem} (d)), and WDS map (Fig. \ref{fig:experiment_wds_sem} (f)) corresponding to the ``Top'' position under ``High LED'' as shown in Fig. \ref{fig:experiment_melt_pool} (c). It is apparent that these images are different from those shown for the ``Top Right'' region of the same melt pool (see Fig. \ref{fig:experiment_melt_pool} (c)), and one can argue whether this structure is cellular or equiaxed dendritic. It is assumed that the tail of the melt pool exhibits a solidification growth rate equal to the laser scan speed (according to Eq. \ref{eq:growth_rate}). We propose that the structure shown at the ``Top'' region under ``High LED'' to be the transverse section of cells growing near the top surface (illustrated as the circular structures in Fig. \ref{fig:cell_illustration} (b)) with a growth rate close to the laser scan speed due to the following reasons. First, the size of the equiaxed dendrites typically are different from each other due to the variation in the nucleation time of each dendrite. However, the similar sizes of the microstructural features that appear in the top region suggests these structures to be cells at steady state condition. Second, according to classical $G$ versus $R$ analysis \cite{kou2003welding}, equiaxed dendrites are formed under low G/R while cells are stable at a higher G/R. Thermal model simulations predicted a higher G/R at the ``Top'' region compared to the ``Top Right'' region ``suggesting'' the structure at the ``Top'' region to be cellular. 
    
    In the ``Low LED'' case, the SEM micrograph (Fig.~\ref{fig:experiment_wds_sem} (g)) corresponding to the ``Top'' region of the melt pool (see (Fig.~\ref{fig:experiment_melt_pool} (a)) reveals a planar structure with a uniform composition as shown in the corresponding WDS map (Fig.~\ref{fig:experiment_wds_sem} (i)).
    
    To compare the composition variations in the aforementioned three regions, line scans (the middle column in Fig.~\ref{fig:experiment_wds_sem}) are extracted from the corresponding WDS maps (the right column in Fig.~\ref{fig:experiment_wds_sem}). The corresponding locations of these line scans are indicated with dashed lines in yellow color in the SEM micrographs (the left column in Fig.~\ref{fig:experiment_wds_sem}). For the cellular structures, the maximum segregation is measured to be Ni-5 at.\% Nb based on 40 different line scans obtained from the WDS maps of different regions in a melt pool. This is also clearly visible in the WDS Line Scans . Local composition in the planar melt pool structure displays small fluctuations in the data attributable to noise, however, the maximum composition is shown to be much lower than the value obtained from that of the cellular structure. In each figure, the alloy composition of Ni-3.2 at.\% Nb is shown (with a dashed line in black color ) as a reference. These results indicate that the microstructure varies locally in a single melt pool (e.g. planar vs. cellular in ``High LED'') under constant process parameters as well as from melt pool to melt pool (e.g. planar in ``Low LED'' vs cellular in ``High LED'') depending on the variation in the applied process parameters.

    \subsection{General Features of the Microstructure}

   %%%%%%%%%%%%%%%%%%%%%%%%
   %%%%%%%%%%%%%%%%%%%%%%%%

      \begin{figure}[ht]
       \centering
       \includegraphics[width=0.9\columnwidth]{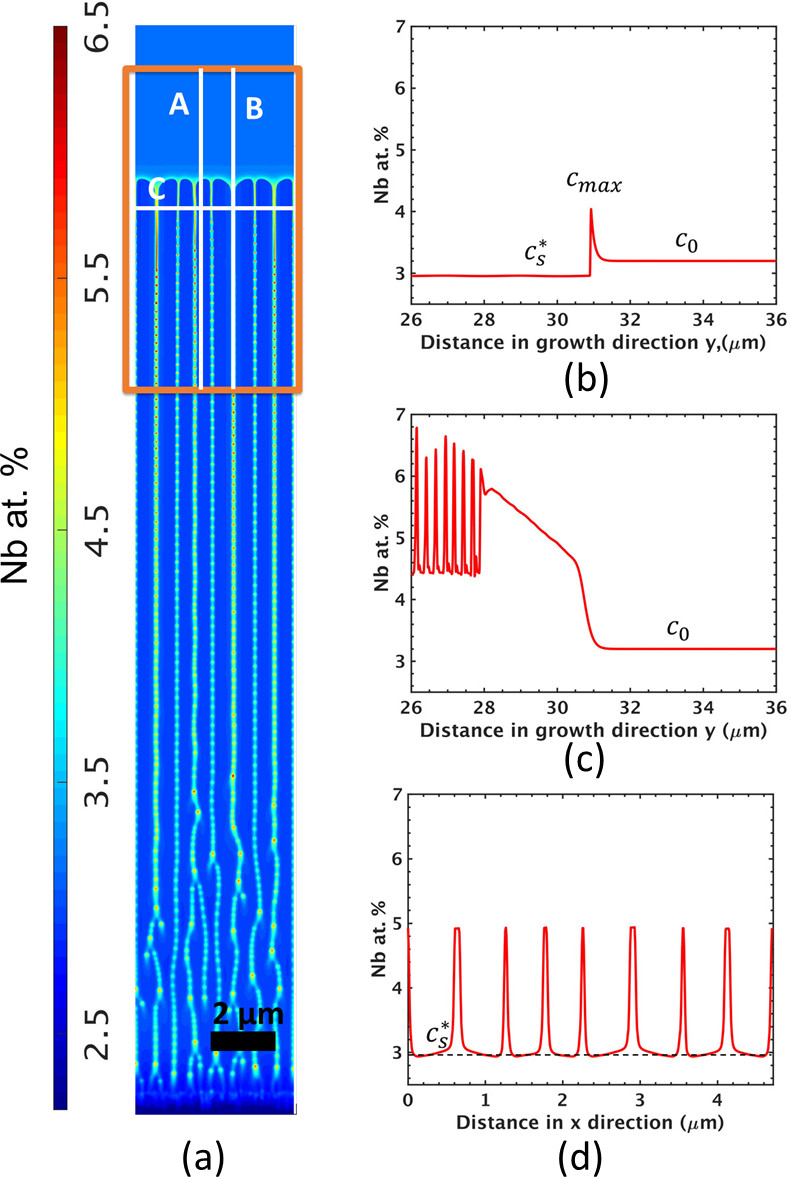}
       \caption{Typical cellular microstructure forms under L-PBF condition. Nb concentration varies along line A (a), B (b), and C (c).}
       \label{fig:typical_micro}
   \end{figure}
   
      \begin{figure}[ht!]
    \centering
    \includegraphics[width=1\columnwidth]{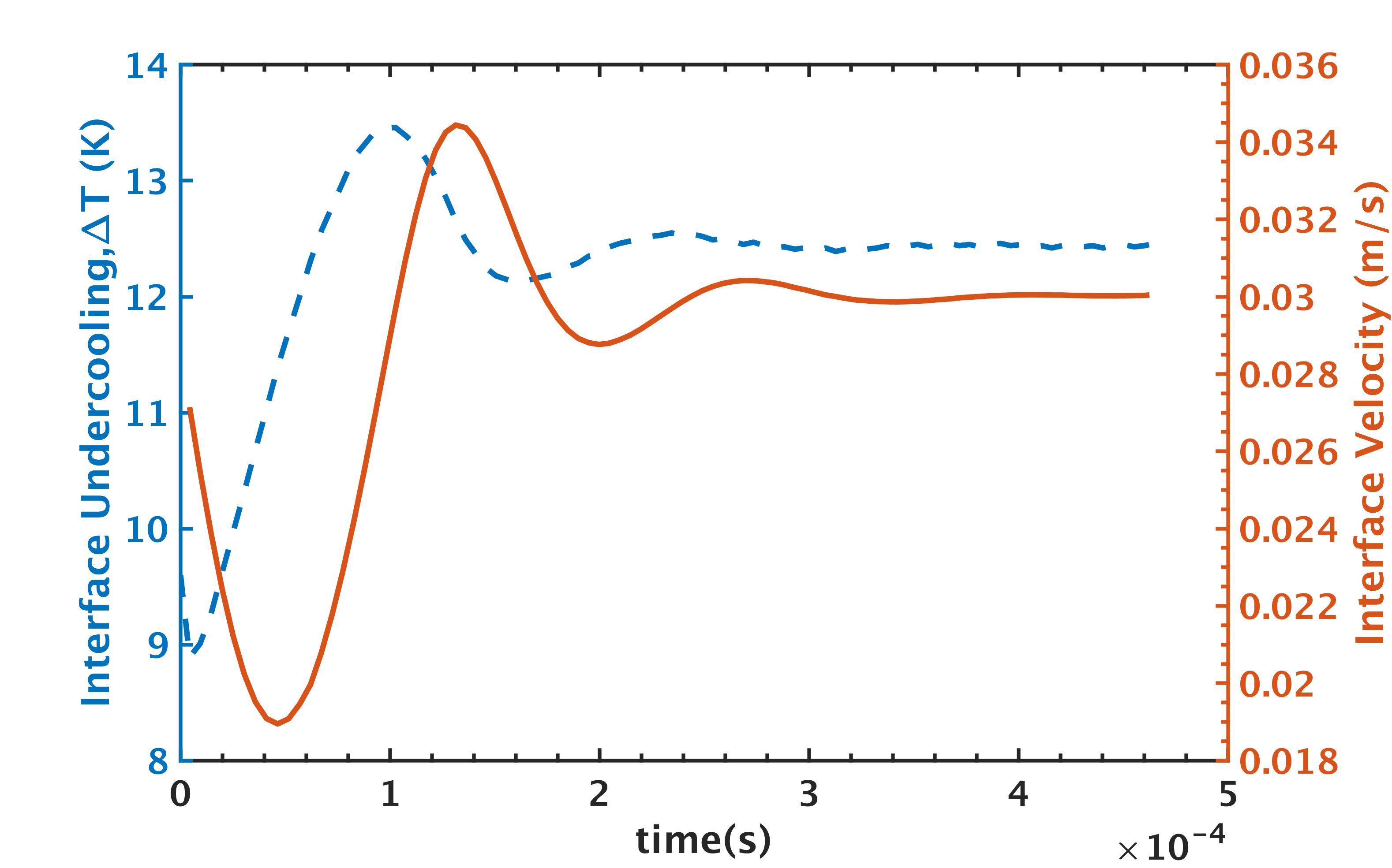}
    \caption{Phase field evolution of the tip velocity and tip undercooling with time.}
    \label{fig:interf_vel_undercool}
    \end{figure} 
    
   %%%%%%%%%%%%%%%%%%%%%%%%
   %%%%%%%%%%%%%%%%%%%%%%%%
    
    Understanding non-planar interfaces (cellular and dendritic) is essential since these segregation structures lead to microsegregation and secondary-phase formation, which affect the  mechanical properties and performance of the solidified material. The characteristic features of a typical non-planar interface (cellular) under AM conditions will be discussed in the following paragraphs.
    
    Figure \ref{fig:typical_micro} (a) represents a typical cellular microstructure predicted in our simulations with the values of $G=3\times10^6 K/m$ and $R=0.03 m/s$. Note that, these values of $G$ and $R$ and in turn the simulated microstructure correspond to a particular location in the melt pool. Variation in the morphology and size of the microstructural features under varying $G$ and $R$ across the melt pool will be discussed in the Section \ref{sec:Micro_Var}. The simulation domain was initialized with a thin layer of solid $FCC-\gamma$ at the bottom and liquid with the composition of Ni-3.2 at.\% Nb on top. The initial solid-liquid interface is perturbed randomly to promote the growth of the cellular structure. A number of small cells appear at the fluctuated interface and grow into the liquid in the direction of applied temperature gradient. As solidification advances, competitive growth at the cell fronts occurs and only a few of these fronts become primary cells. The primary dendrite arm spacing (PDAS) varies at the initial stages and remains constant once the steady state is reached. At this stage, the cell tips advance at a constant velocity and constant temperature as shown in Fig. \ref{fig:interf_vel_undercool}. Note that the interface velocity reaches the the applied solidification growth rate $R=0.03m/s$ at steady state.
    
    Solute microsegregation is a typical phenomenon observed in the rapid solidification process and has significant influence on the mechanical properties of the solidified material. As cells grow, Nb is rejected into the liquid, resulting in Nb enriched intercellular regions as shown in Fig. \ref{fig:typical_micro} (a). The Nb enriched liquid droplets appearing along the cell grooves as solidification advances are also shown in Fig. \ref{fig:typical_micro}. With time, these Nb enriched droplets may transform to secondary phases such as $Ni_3Nb$ if the amount of microsegregation is enough to promote the nucleation and growth of this phase. The Scheil simulations~\cite{keller2017application} indicated that for the selected alloy composition (Ni-3.2 at.\% Nb), at least Ni-7 at.\% Nb is needed for such a transformation to occur. However, the Nb amount in those droplets have not reached the levels needed for the precipitation of the secondary phases.

    In order to investigate solute segregation, three lines denoted by $A$, $B$, $C$ are selected in Fig. \ref{fig:typical_micro} (a) and the corresponding Nb concentration profiles along the selected lines are presented in Fig. \ref{fig:typical_micro}(b) to \ref{fig:typical_micro}(d). Figure \ref{fig:typical_micro}(b) demonstrates the Nb concentration variations through the core of the cell into the liquid along the growth direction $y$ (line A in Fig. \ref{fig:typical_micro} (a)). The left side of this profile denoted by $c_s^{*}$ corresponds to the concentration in the cell core near the solid-liquid interface (cell tip), while the spike denoted by $c_{max}$ represents the concentration at the liquid side of the interface. Beyond this, the concentration of Nb decreases rapidly and eventually reaches the far-field liquid composition given by $c_0$. 
    
    In Fig. \ref{fig:typical_micro} (c), the Nb concentration variation is shown along the intercellular region (line B in Fig. \ref{fig:typical_micro} (a)). Nb concentration gradually decreases in the growth direction $y$ along the intercellular region and reaches the far-field liquid concentration of Ni-3.2 at.\% Nb beyond the cell tips. The slope of the linearly decaying part, between $28 \mu m$ and $30 \mu m$, in Fig. \ref{fig:typical_micro}(c) is calculated to be 3.33$\%/\mu$m. An analytical solution to obtain the solute concentration gradient in the intercellular region was reported by \cite{bower1966measurements}. Ignoring the interface curvature effects, the concentration gradient in the above region can be estimated by $dc/dx = G/m$, where $G$ is the applied temperature gradient,$m$ is the liquidus slope. $G$ and $m$ in this work are $3\times10^6K/m$ and $-869 K/\%$, respectively. Substituting the values in the equation, the solute concentration gradient is found to be 3.45$\%/\mu$m, which agrees well with the calculated value.
    
    In Fig. \ref{fig:typical_micro}(d), the Nb concentration variation is presented perpendicular to the growth direction (line C in Fig. \ref{fig:typical_micro} (a)). Here, the top and bottom of the U-shaped profile correspond to the Nb concentration in the intercellular regions and the cell cores, respectively. Note that $c_s^{*}$ in Fig. \ref{fig:typical_micro}(b) corresponds to the bottom of the U-shaped profile, representing the concentration at the cell core. It is observed that the concentration at the intercellular region is much higher than the concentration inside the cell cores. This happens due to the fact that the solute is rejected from the growing cells into the liquid and since the growth of cells happens rapidly, there is no time for the solutes to diffuse back and eventually enrichment of Nb occurs.

    \subsection{Microstructural Variability as a Function of Thermal Parameters ($G$, $R$)}\label{sec:Micro_Var}

    \begin{figure}[!t]
     \centering
     \includegraphics[width=1\columnwidth]{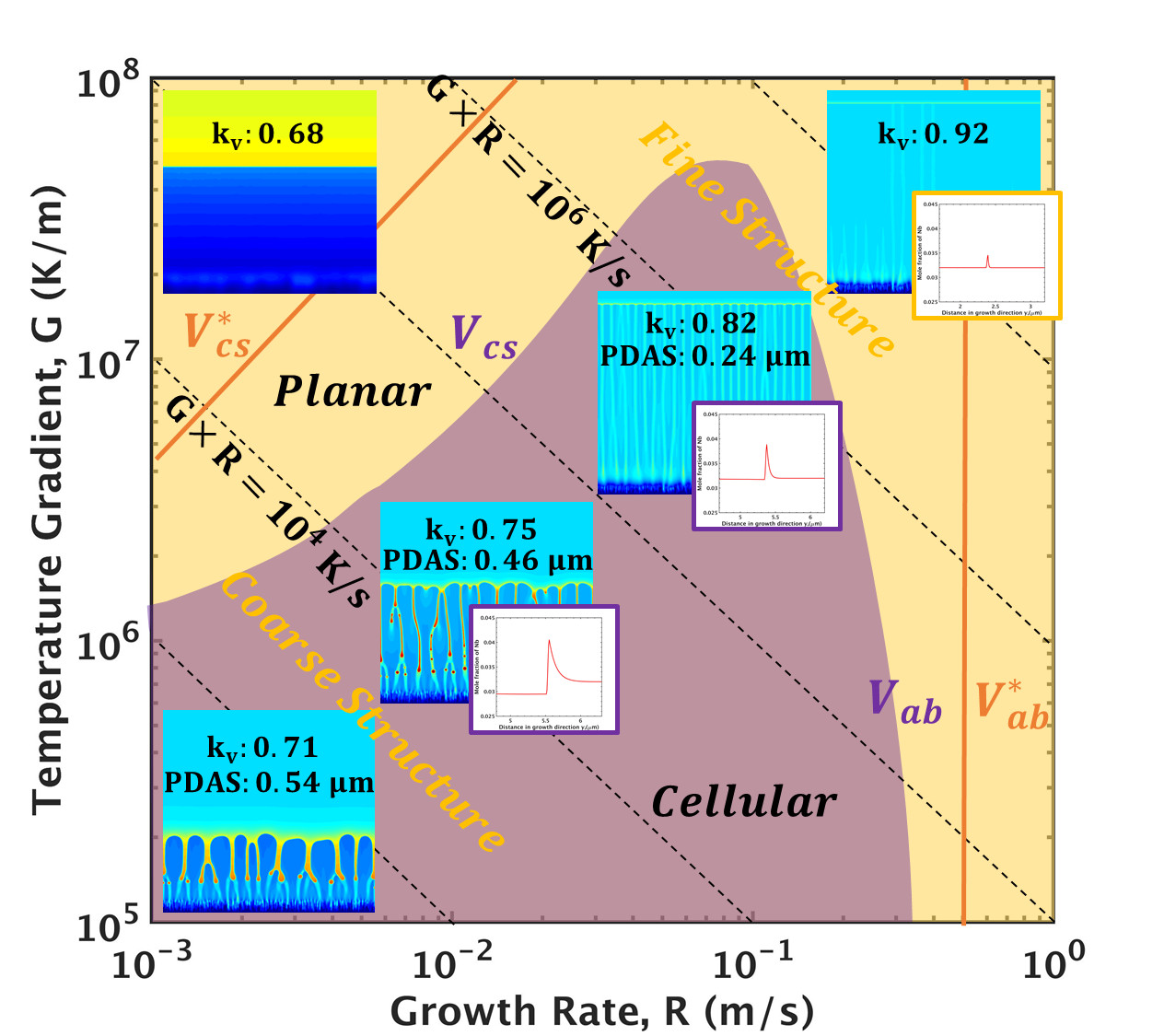}
     \caption{The variation in morphology, size of cellular segregation structure , and microsegregation as a function of growth rate and temperature gradient is shown.}
     \label{fig:GV_map}
    \end{figure}

    % UPDATED FIGURE %
    \begin{figure*}[!hb]
     \centering
       %\subfloat[]{\includegraphics[width=0.5\columnwidth]{images/new/pdas_vs_cooling.jpeg}}
        %\subfloat[]{\includegraphics[width=0.5\textwidth]{images/after_revision/pdas_vs_cooling_with_exp_paper_v2.jpeg}}
        \subfloat[]{\includegraphics[width=0.5\textwidth]{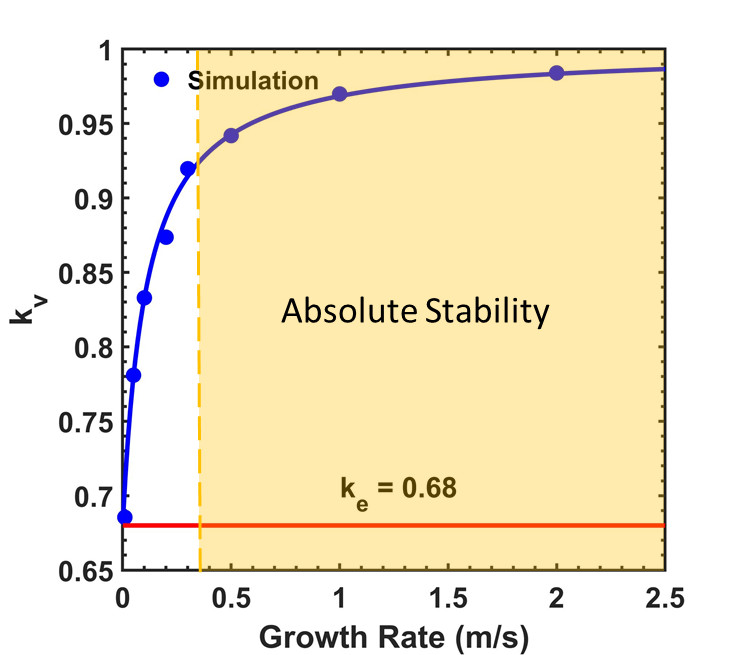}}
        \subfloat[]{\includegraphics[width=0.5\textwidth]{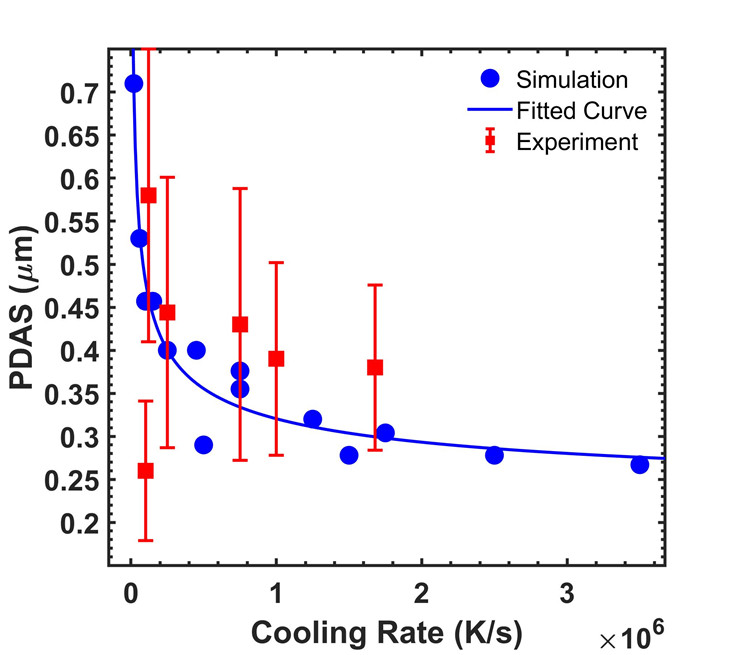}}      
     \caption{The calculated primary dendrite arm spacing (PDAS) reduces as the cooling rate increases.As the solidification growth rate $R$ increases the segregation coefficient increases. }
     %\hl{~(UPDATED FIGURE)}
     \label{fig:kv_vs_speed} 
     % \label{fig:PDAS_vs_cooling}
    \end{figure*}

    % UPDATED FIGURE %
    \begin{figure*}[!ht]
      \centering
      \subfloat[]{\includegraphics[width=0.5\textwidth]{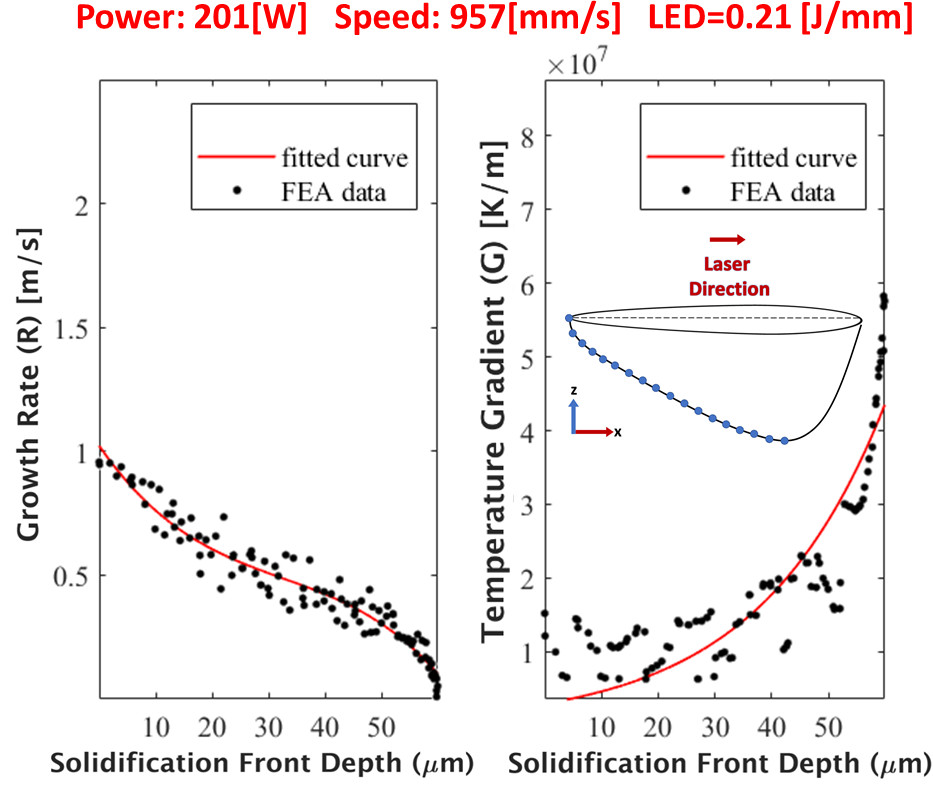}}
      \subfloat[]{\includegraphics[width=0.5\textwidth]{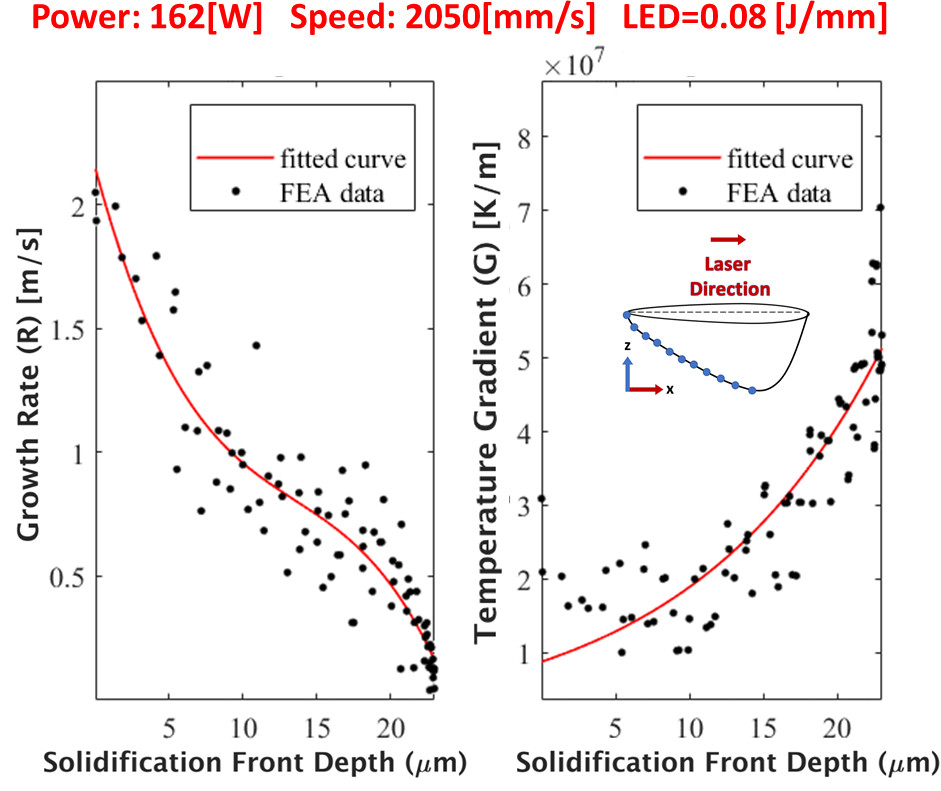}}\\
      \caption{The predicted temperature gradient, $G$ and growth rate, $R$ along the centerline of melt pool boundary are shown for varying solidification front depths for two set of process parameters.The melt pool schematics with blue dots demonstrating the representative locations of $G$ and $R$ predictions are also included for each case.}
      \label{fig:GR_longi}
    \end{figure*}

    % UPDATED FIGURES %
    \begin{figure*}[!ht]
      \centering
     \includegraphics[width=0.8\textwidth]{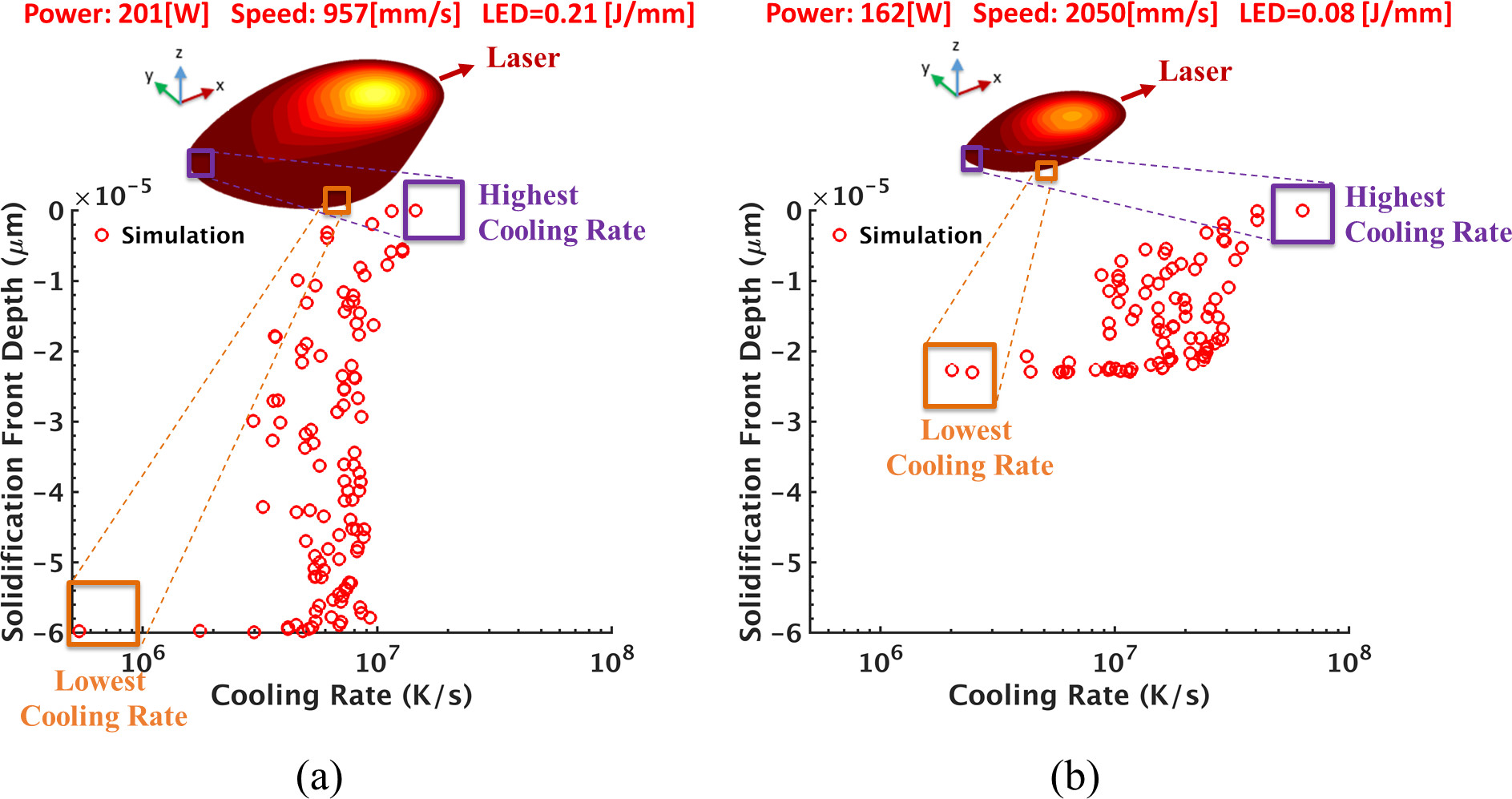}
      \raisebox{0.2cm}{\includegraphics[width=0.8\textwidth]{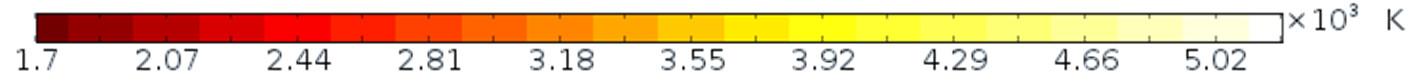}}
      \caption{The predicted cooling rates at varying solidification front depths are shown for two set of process parameters.The oblique view of predicted melt pool geometries along with the locations of highest and lowest cooling rates throughout the centerline of melt pools are also demonstrated. The red arrows located on the front side of the melt pools point the direction of the moving laser heat source for each case.}
      \label{fig:cooling_vs_depth}
    \end{figure*}  
    
    The temperature gradient, $G$, and the solidification growth rate, $R$, are the most significant parameters in determining the solidification microstructure. The ratio of these parameters, $G/R$, determines the morphology of the solidification microstructure (e.g. planar, cellular, columnar dendritic, and equiaxed dendritic), while the product of these parameters $G\times R$, the cooling rate $\dot T$, determines the size of the microstructure (the higher the cooling rate the finer the structure is). 
    
    \begin{figure*}[ht!]
     \centering
     \includegraphics[width=0.85\textwidth]{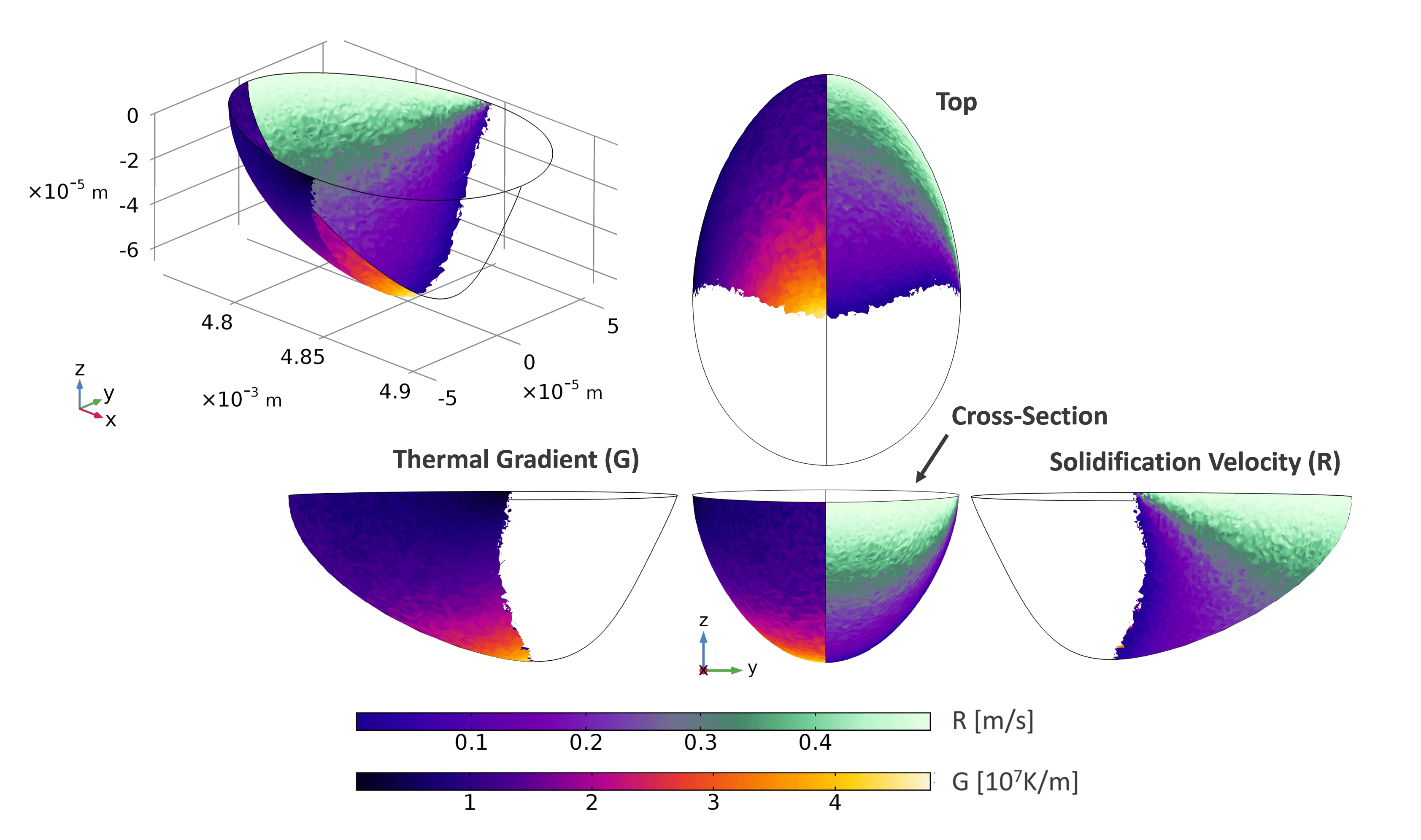}
     \caption{Variation in the $G$ and $R$ are shown across the melt pool.}
     \label{fig:3d_melt_pool}
    \end{figure*} 
    
    A solidification morphology selection map for Ni-3.2 at.\% Nb demonstrating the variation in the morphology and size of cellular segregation structure as a function of $G$ and $R$ is illustrated in Fig. \ref{fig:GV_map}. For the selected alloy composition under the L-PBF process with varying $P$ and $\vec{v}$, two types of growth morphologies, planar and cellular, are predicted.   As the growth rate increases, a transition from planar to cellular and again to planar morphology is predicted. Typically, the constitutional supercooling criterion~\cite{kurz1989fundamentals} and the absolute stability criterion~\cite{kurz1989fundamentals} are utilized to roughly estimate these limits of growth rates in between which the cellular morphology is stable, given by $V_{cs} < R < V_{ab}$. Here, the lower limit is determined by the constitutional supercooling criterion, $V_{cs}=\frac{G D_L}{\Delta T}$, while the upper limit is approximated by the absolute stability criterion, $V_{ab}=\frac{\Delta T D_L}{ke \Gamma}$, where $\Delta T$, $ke$, $\Gamma$ are the equilibrium freezing range, equilibrium segregation coefficient, and Gibbs-Thomson coefficient, respectively. Beyond these limits, a planar structure becomes stable (either for low growth rates ($R < V_{cs}$) or for extremely high growth rates ($R > V_{ab}$)).
    
    The aforementioned solidification limits are determined in two ways: the classical theory (linear solid lines) and PF simulations. The theoretically determined limits are indicated with $V_{cs}^{\star}$ and $V_{ab}^{\star}$, while the limits predicted by PF simulations are indicated with $V_{cs}$ and $V_{ab}$ as shown in Fig. \ref{fig:GV_map}. 
    Note that the boundaries for $V_{cs}$ and $V_{ab}$ are established based on $\approx$ $70$ simulations at different $G$ and $R$ conditions. The predicted growth morphologies (e.g. cellular vs planar) for each $G-R$ combination were marked in the solidification map. The planar/cellular boundary is then roughly determined.
    A triangular region in the middle of the map is indicated as cellular in both cases. Less than an order of magnitude difference is shown between the lower and upper limits predicted from model and theory especially for high $G$ conditions, while much better agreement is achieved as $G$ decreases. Note that the theory gives a rough estimate of the limits and has some limitations. For example, the constitutional supercooling criterion ignores the effect of surface tension, which is considered in our PF simulations. Also, the absolute stability criterion has other limitations. For instance, it accounts only for the species diffusion in the liquid and thus neglects solid diffusion, which is taken into account in PF predictions. Therefore, we believe that the theoretical predictions are less accurate, especially under rapid solidification conditions.

    In addition to the type of growth structures (planar and cellular), the presented solidification map in Fig. \ref{fig:GV_map} can be utilized to gain information on the variation in size of cells (PDAS) and microsegregation ($k_v$) as a function of $G$ and $R$. 

    Note that the PDAS predictions reported in the present work are their average values, calculated by dividing the number of cells at the steady state to the width of the simulation domain, similar to those reported in the previous research efforts \cite{ghosh2017primary}.  
    
    As the cooling rate increases (from the bottom left corner to the top right corner), PDAS is reduced (within $V_{cs}<R<_{Vab}$) and finally the interface re-stabilizes and a segregation-free planar structure is achieved  (beyond $R>V_{ab}$). In the majority of current literature, a cellular/dendritic structure with nonuniform properties is predicted/observed as the typical growth structure under AM \cite{ghosh2017primary,acharya2017prediction,kundin2019microstructure}. We emphasize the possibility of a planar structure with uniform properties at high cooling rates ($~10^7-10^8 K/s$).

    Increasing the cooling rate or growth rate results in a decrease in the calculated $k_v$. Under equilibrium conditions in the selected alloy $k_v$ can be calculated as $k_e=0.68$. Our simulations demonstrated that $k_v=k_e$ with $c_s^0=c_s^{eq}$ is achieved only for the low growth rates, observed near the fusion boundary, and resulted in a planar growth structure as demonstrated in top left image in Fig. \ref{fig:GV_map}.  This finding agrees well with the Nb depleted region near the fusion boundary as presented in Fig. \ref{fig:experiment_wds_sem}. Beyond the fusion boundary  and through the center line of the track, the growth rate is much higher and the resultant segregation coefficient given by $k_v$ diverges from the equilibrium value and eventually reaches $k_v~=1$, indicating the full solute trapping condition, which takes place during rapid solidification beyond the velocity limit of absolute stability ($R>>V_{ab}$) as displayed in the top right image in Fig. \ref{fig:GV_map}. The inset images in Fig. \ref{fig:GV_map} illustrate the variation in the concentration profile (in mole fraction Nb) and the calculated $k_v$ for the representative microstructure predictions. It is shown that as cooling rate/growth rate increases, the peak of the profile reduces while the concentration at the left and right side of the profile approach and become equal with the value of $c_0$ and $k_v~=1$, indicating solute trapping. It should be noted that the finite interface dissipation PF model utilized is well suited to study the rapid solidification phenomenon such as solute trapping. By adjusting the permeability parameter $P^{intf}$ (Eq.~\ref{eq:permeability}), we can quantify the solute trapping behavior, as well as the non-equilibrium segregation coefficient. The influence of permeability on calculated $kv$ will be shown in Section \ref{effect_of_perm}.

    \begin{figure*}[!ht]
     \centering
     \subfloat[]{\includegraphics[width=0.3\textwidth]{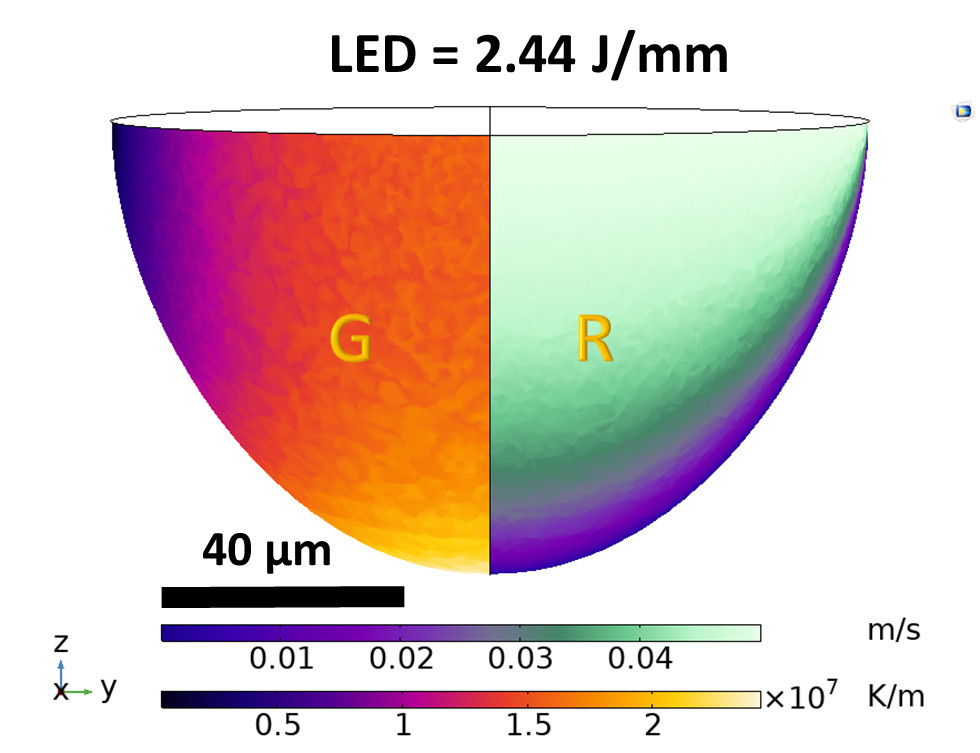}}
     \subfloat[]{\includegraphics[width=0.3\textwidth]{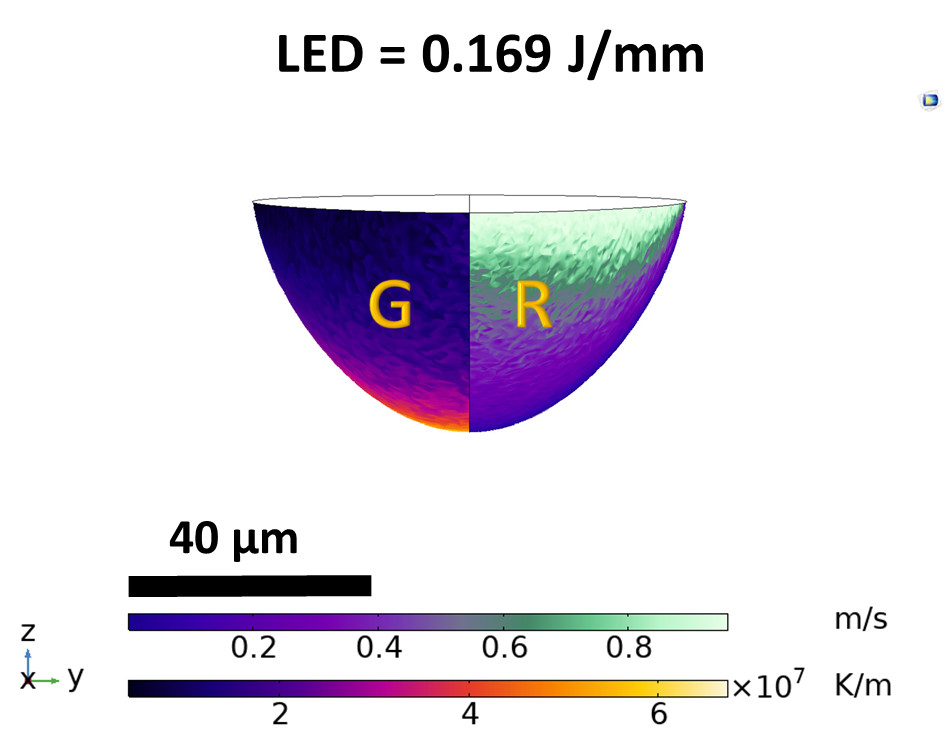}}
     \caption{The calculated G and R along the transverse section of the melt pool boundary are shown for two set of process parameters with different linear energy densities (LED).}
     \label{fig:transverse_melt_pool_GR}
    \end{figure*}
    
    The variation in PDAS and $k_v$ as a function of $G$ and $R$ are further analyzed in Fig. \ref{fig:kv_vs_speed}.  As discussed in the above paragraphs, during rapid solidification the system deviates from local equilibrium. To quantify this deviation, the velocity-dependent partition coefficient $k_v$, given by: $k_v(R)=c_s^*/c_{max}$ \cite{danilov2006phase} is plotted against the growth rate as shown in Fig. \ref{fig:kv_vs_speed}(a). The horizontal line in Fig. \ref{fig:kv_vs_speed}(a) represents the equilibrium segregation coefficient of the NiNb alloy. As the solidification growth rate $R$ increases, the calculated segregation coefficient deviates from this value and approaches to 1 beyond the velocity limit of absolute stability indicated with a vertical dashed line in Fig. \ref{fig:kv_vs_speed}(a). Above this limit, the solid-liquid interface restabilizes to a planar interface.
    
    Fig. \ref{fig:kv_vs_speed}(b) presents the variation in predicted PDAS with respect to the cooling rate ($G\times R$) along with a fitted line and experimental measurements obtained from the regions with cellular structure as presented in Fig. \ref{fig:mid_led_valid} and \ref{fig:high_led_valid}.  The theoretical analysis \cite{hunt1979cellular,kurz1981dendrite} suggests that the PDAS decreases as the cooling rate ($G\times R$) increases. In the present work, the simulation results reveal a clear trend and agree with the theory. A similar trend is also observed for the experimentally measured PDAS excluding the first data point with an average PDAS of $0.26 \mu m$. 
    
    It is worth mentioning that the first two experimental data points in Fig. \ref{fig:kv_vs_speed}(b) were taken from the same region of a single melt pool (from the top right in high LED melt pool as shown in Fig. \ref{fig:high_led_valid}) with slightly different locations. The FE model predicts similar cooling rates for each location. Therefore, similar PDAS are expected. However, the measured PDAS for these two locations were found to be clearly different (i.e. $\approx 0.26 \mu m$ and $\approx 0.58 \mu m$), contrary to expectations. Among all experimental data points presented in Fig. \ref{fig:kv_vs_speed}(b), these two points are anticipated to show the highest PDAS measurements due to the lowest cooling rates predicted by FE model. The measurement of $0.58 \mu m$ corresponds more closely to our microstructural predictions and expectations based on the predicted $G \times R$ value in the region compared to the value of $0.26 \mu m$, hence the latter is believed to be an outlier. Considering the uncertainties associated with the experiments as well as the model, it is not straightforward to explain the main reason of the outlier. This could be due to the simplified FE model assumptions (e.g. neglecting fluid flow), or uncertainties in the experiments (e.g. the possible local fluctuations in temperature gradient, unknown orientation of grains, etc.).
    
    The typical cooling rate under AM ranges from $10^5 K/s$ to $10^8 K/s$ depending on the process parameters and the location in the melt pool. The predicted PDAS varies from $0.58 \mu m$ to $0.2 \mu m$, as the cooling rate increases from $10^5 K/s$ to $3.5\times10^6 K/s$. Above this, a transition from cellular to planar interface is observed. Note that, although both increasing $G$ and $R$ leads to an increase in the cooling rate ($G \times R$), hence smaller PDAS, their effects might be different. For example, at $G=5\times10^5 K/m$ and $R=3\times10^{-1} m/s$ fine cellular structure is observed with the PDAS of $0.3 \mu m$ while a planar structure for $R < V_{cs}$ is observed at $G=1.5\times10^7 K/m$ and $R=1\times10^{-2} m/s$. Although in both cases the cooling rate is calculated as $G \times R=1.5\times10^5 K/s$, two different growth structures are observed. In another example with the condition of $G=5\times10^5 K/m$ and $R=1\times10^{-1} m/s$ results in a PDAS of $0.4 \mu m$, whereas $G=5\times10^6 K/m$ and $R=1\times10^{-2} m/s$ leads to a PDAS of $1 \mu m$. Therefore, one should be careful using PDAS vs. cooling rate information when there are a few orders of magnitude variation in $G$. On the other hand, by keeping the $G$ constant and varying $R$, a consistent correlation between PDAS and cooling rate can be obtained. Note that, Fig. \ref{fig:kv_vs_speed} (b) is created by varying $G$ within the range of $10^6-10^7 K/m$  and $R$ from $R=10^{-2} m/s$ to $R=10^{-1} m/s$. Therefore, the information provided in this plot will be valid only for the given ranges of $G$ and $R$.

    \subsection{Microstructural Variability as a  Function of Process Parameters}

    Microstructural features vary spatially within a single melt pool as well as from melt pool to melt pool depending on the process parameters applied. It is therefore essential to understand and control these variabilities by tailoring the process parameters so that a final product with desired properties can be achieved. In this subsection, the influence of process parameters on the thermal parameters ($G$, $R$) will be discussed first, followed by a description of the microstructural features as a function of these estimated thermal parameters.

    % UPDATED FIGURE %
    \begin{figure}[!htb]
     \centering
     \includegraphics[width=1\columnwidth]{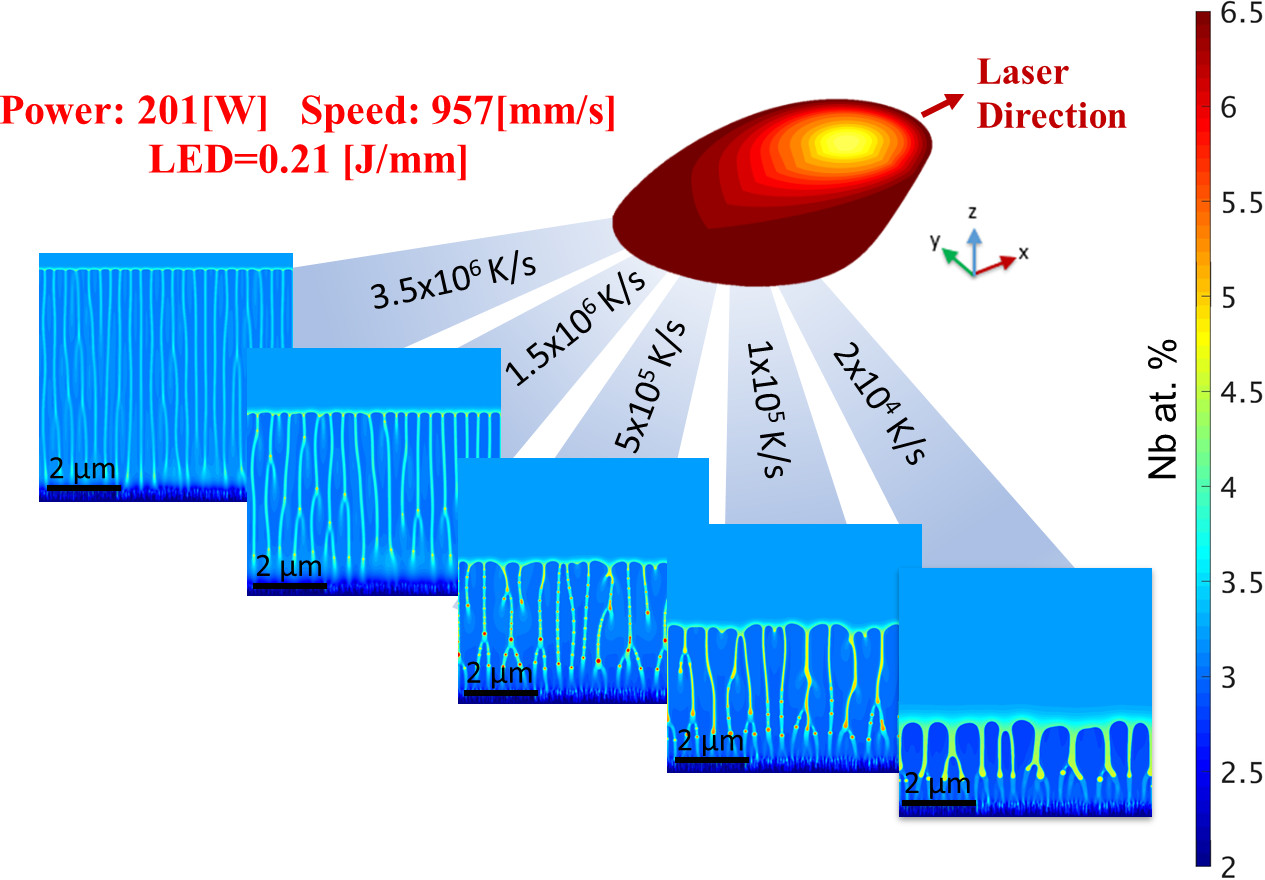}
     \caption{Variation in the cellular structure and size at varying cooling rates along the melt pool boundary is shown for high LED condition.}
     \label{fig:high_led}
    \end{figure}

    % UPDATED FIGURE %
    \begin{figure}[!ht]
     \centering
     \includegraphics[width=0.9\columnwidth]{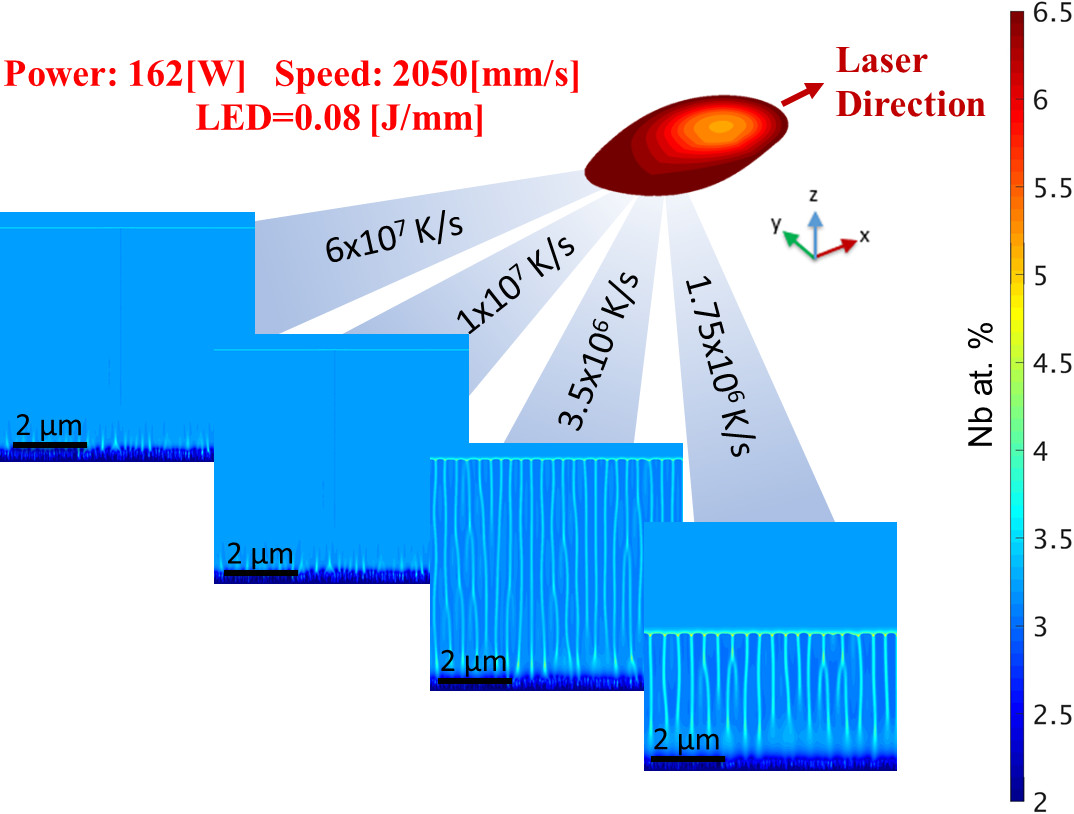}
     \caption{Variation in the solidification structure and size at varying cooling rates along the melt pool boundary is shown for low LED condition. A transition from cell to plane structure is observed at high cooling rates.}
     \label{fig:low_led}
    \end{figure}

    \subsubsection{Effects of Temperature Gradient, Growth Rate, and Cooling Rate}
   
    Single track laser melting simulations were run at varying laser power ($P: 70-255 [W]$) and laser speed ($\vec{v}: 50-2300 [mm/s]$). To investigate  spatial variations in the microstructural features, the output $G$ and $R$ were predicted both along the longitudinal section and transverse section of the resultant melt pool. Fig.~\ref{fig:GR_longi} represents an example of output $G$ and $R$ for two sets of $P$ and $\vec{v}$ along the longitudinal section of the melt pool boundary, corresponding to high and low LED cases. The melt pool schematics with blue dots demonstrating the representative locations of $G$ and $R$ predictions along the centerline of melt pool boundary are also included for each case. Depending on the location of the melt pool, the calculated G and R varied within the range of [$3\times10^6 K/m - 5\times10^7 K/m$] and [$3\times10^{-5} m/s - 1 m/s$], respectively. The maximum $R$ is calculated near the top of the melt pool (with a low solidification front depth) while the minimum is found near the bottom of the melt pool (with a high solidification front depth). In contrast, the maximum $G$ is calculated at the bottom of the melt pool as the minimum $G$ is observed at the top.

%%%%%%%%%%%%%%%%%%%%%%%%%%%%%%%%%%%
    
    \begin{figure}[!ht]
     \centering
     \includegraphics[width=0.8\columnwidth]{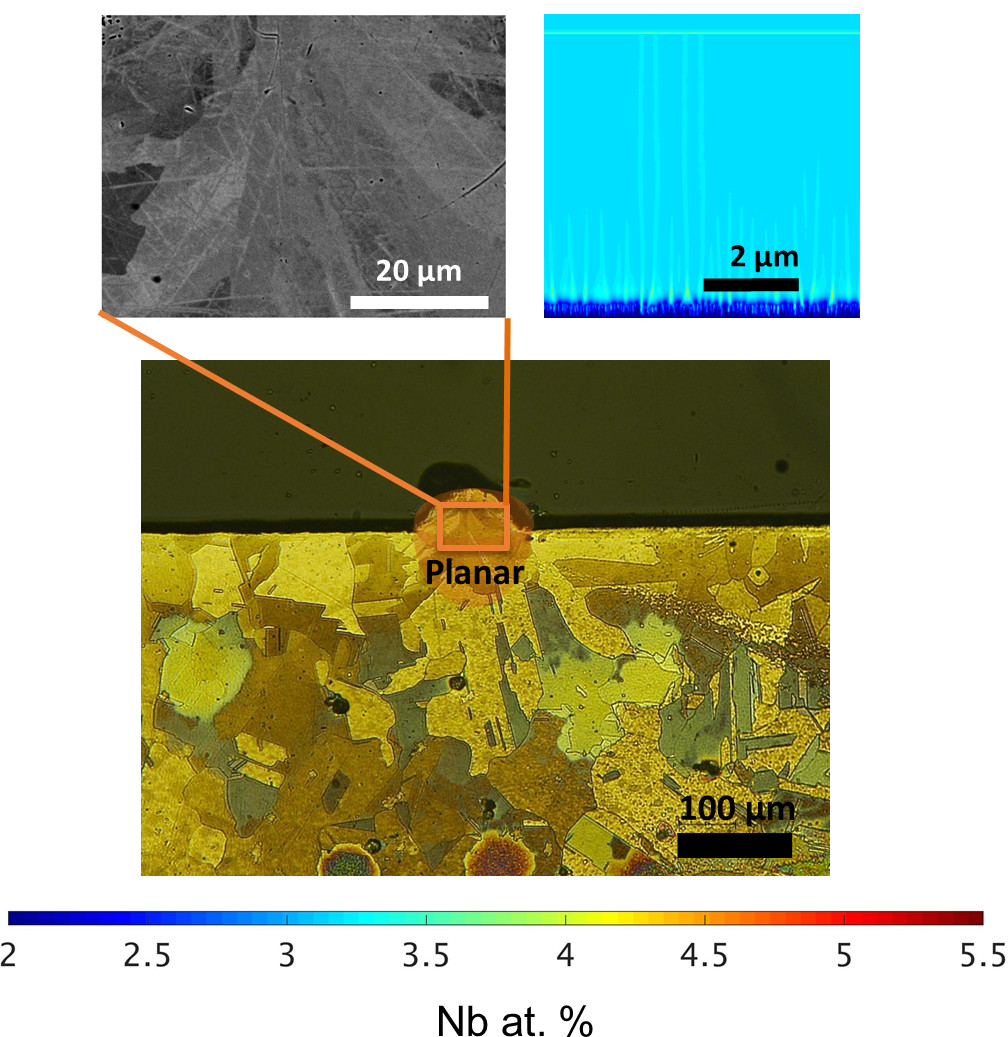}
     \caption{The predicted growth morphology and size throughout the melt pool are demonstrated along with the experimental measurements. Planar structure is observed throughout the melt pool. ($P$: 162 W,  $\vec{v}$: 957 mm/s, LED: 0.169 J/mm)}
     \label{fig:low_led_valid}
    \end{figure}

    Due to the inverse relationship between solidification front depth and $R$, and the proportionality between front depth and $G$, it is not clear from Fig. \ref{fig:GR_longi} how the resultant cooling rate $\dot T = G\times R$  will vary as a function of the depth. In figure \ref{fig:cooling_vs_depth}, the variation of $\dot T$ along with the centerline of melt pool boundary with increasing depth is explicitly shown for the same process parameters as in Fig. \ref{fig:GR_longi}. The oblique view of predicted melt pool geometries along with the locations of highest and lowest cooling rates throughout the centerline of melt pools are also demonstrated. The red arrows located on the front side of the melt pools point the direction of the moving laser heat source for each case. On the left, an example of a high LED case is shown, while on the right a low LED case is presented. In both cases, the cooling rate increases as we move from the bottom to the top along the melt pool boundary. Predicted cooling rate varies as a function of LED. Lower LED tends to show higher rates of cooling. While the maximum predicted cooling rate in the high LED case is around $\dot T=1.5\times10^7K/s$, the maximum rate of cooling for low LED is $\dot T=7\times10^7K/s$. Variations in the cooling rate affect the solidification microstructure, which will be discussed in the following sections.

%%%%%%%%%%%%%%%%%%%%%%%%%%%%%%%%%%%%%    

    Fig. \ref{fig:3d_melt_pool} presents a top, side (longitudinal), cross-section (transverse), and oblique views of the three dimensional melt pool at $P$:162 [W] and $\vec{v}$: 500 [mm/s], demonstrating the local variation in the $G$ and $R$ throughout the melt pool boundary. The same trend for $G$ and $R$ as presented in Fig. \ref{fig:GR_longi} is shown along the centerline of melt pool boundary in the longitudinal view (as the depth increases along the melt pool boundary, $G$ increases while $R$ decreases). Note that, the longitudinal view in Fig. \ref{fig:3d_melt_pool} provides the predictions of $G$ and $R$ along the entire melt pool boundary (including both centerline and points away from the centerline).It can be inferred that while $R$ shows a clear dependence on both lateral position and depth, $G$ has a strong dependence on depth only.

    % UPDATED FIGURE %
    \begin{figure}[!ht]
     \centering
     \includegraphics[width=1\columnwidth]{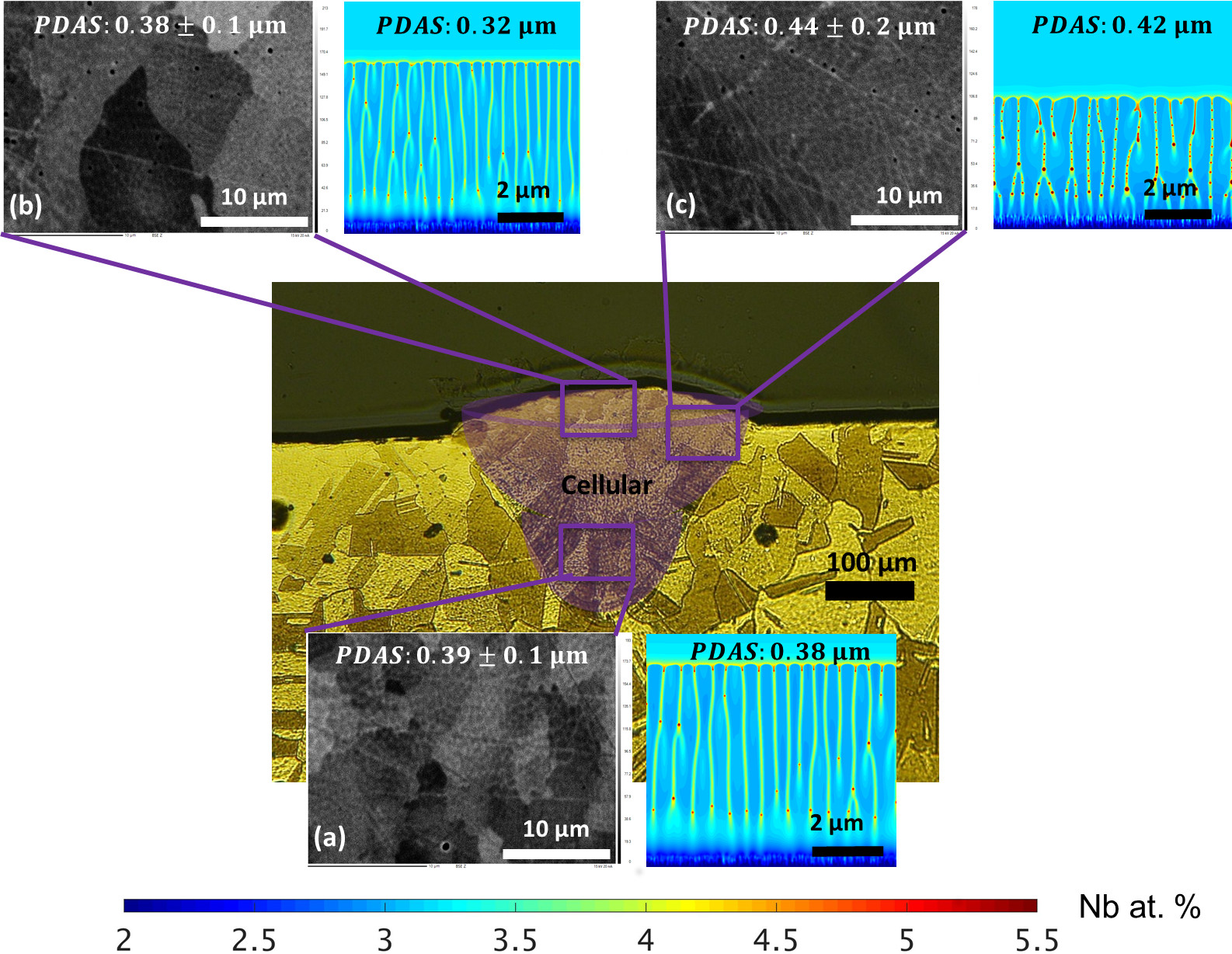}
     \caption{The predicted growth morphology and size throughout the melt pool are demonstrated along with the experimental measurements. Cellular structure is observed throughout the melt pool. ($P$: 96 W,  $\vec{v}$: 67 mm/s, LED: 1.43 J/mm)}
     \label{fig:mid_led_valid}
    \end{figure}
    
    In addition, the local variation in $G$ and $R$ across the melt pool can be seen in the left and right half of the cross-sectional view in Fig. \ref{fig:3d_melt_pool}, respectively. $G$ increases from top ($1.5\times10^6 K/m$) to bottom ($4.8\times10^7 K/m$) across the melt pool with the minimum and the maximum being at the top (left) edge and the bottom of the melt pool, respectively. In contrast, $R$ reduces from the top to the bottom as well as from the interior of the melt pool to the edge of the melt pool.

    In the present work, an experimental analysis of the microstructure is performed at the transverse cross-sections of the melt pool. Therefore, the $G$ and $R$ data across the transverse cross-section will be utilized for the validation purposes. To predict the spatial variation in the microstructural features across the transverse section, $G$ and $R$ data at corresponding locations are extracted from the associated melt pools (Fig. \ref{fig:transverse_melt_pool_GR}). For the experimental validation, we will be using three different cases: low, medium and high LED. Here, we show the variation in $G$ and $R$ for the low and high LED cases as an example. The $G$ and $R$ data needed for the medium LED case is also extracted from the associated melt pool, but not explicitly shown here. The high LED case results in a larger melt pool size with lower $G$ and $R$ values, when compared to the low LED case as expected. This is due to the fact that as the laser speed $\vec{v}$ increases, the heat accumulated in the laser interaction zone decreases, leading to formation of smaller melt pools. In both cases, $G$ increases from the top to the bottom. While a high degree of variation in $G$ from the top-edge to the top-interior location in the melt pool is observed in the high LED condition, it is almost constant in the low LED condition. In contrast to $G$, $R$ decreases from the top to the bottom and from interior to edge in both cases. Since $R$ is a function of $\vec{v}$, it is reasonable to observe higher cooling rate for higher $\vec{v}$. This is the case with the low LED condition.
    
    % UPDATED FIGURE %
    \begin{figure}[!ht]
     \centering
     \includegraphics[width=1\columnwidth]{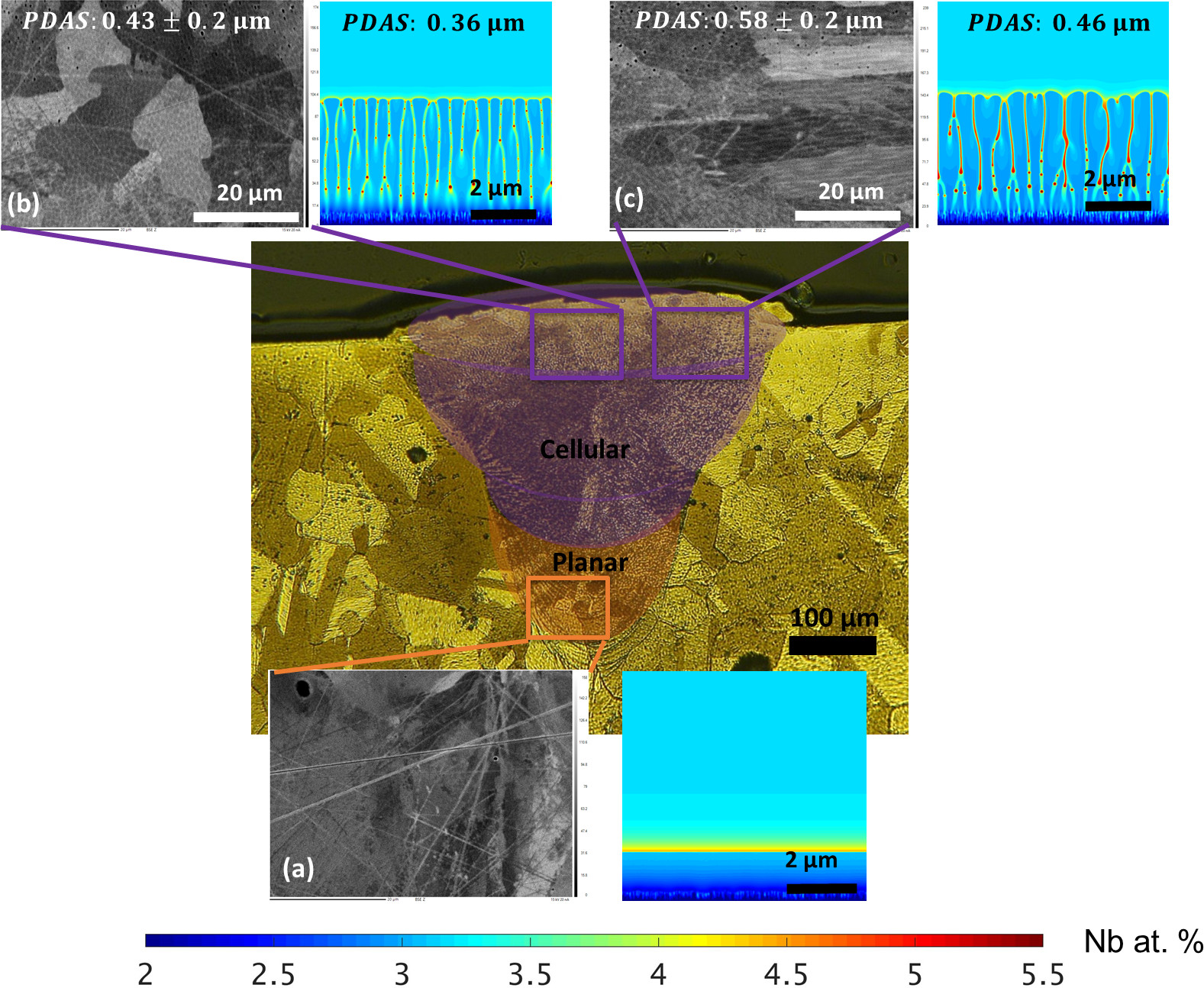}
     \caption{The predicted growth morphology and size throughout the melt pool are demonstrated along with the experimental measurements at the corresponding locations. A transition from a planar to a cellular structure from the bottom to the top of the melt pool is observed. ($P$: 122 W, $\vec{v}$: 50 mm/s, LED: 2.44 J/mm) }
     \label{fig:high_led_valid}
    \end{figure}    
    
    \subsubsection{Morphology and Size}
   
    As mentioned previously, the cellular segregation structure and size vary locally within a melt pool as well as from melt pool to melt pool under different processing conditions. To investigate the variation in the microstructure along the longitudinal section of melt pools under different AM conditions, the calculated cooling rates along the melt pool boundaries, as shown in Fig.~\ref{fig:cooling_vs_depth}, are utilized. 

    Fig.~\ref{fig:high_led} demonstrates the variation in the solidification microstructure at different locations of the melt pool boundary under the high LED condition. It is observed that the resultant cooling rates in this case lead to the formation of cellular structure with varying cell sizes (PDAS) along the melt pool boundary. As the cooling rate increases from the bottom to the top of the melt pool, the cells become finer. Here, the predicted cooling rate varied from $2\times10^4 K/s$ to $3.5\times10^6 K/s$ as we go from the bottom to the top of the melt pool, and the predicted PDAS ranged from $0.7 \mu m$ to $0.3 \mu m$, respectively.

     Fig.~\ref{fig:low_led} demonstrates the variation in the solidification microstructure at different locations of the melt pool boundary under the low LED condition. Here, the predicted cooling rate varied from $1.75\times10^6 K/s$ to $6\times10^7 K/s$ from bottom to top of the melt pool. Near the bottom of the melt pool, a cellular structure is predicted, while a planar structure is observed near the top of the melt pool due to extremely high cooling rates. Note that the transition from cellular to planar structure at high solidification growth rate (hence the cooling rate) is already reported in the literature \cite{ludwig1996morphological}. One can refer to the solidification map in Fig. \ref{fig:GV_map} to gain an understanding of the effect of $G$ and $R$ on such a transition in the growth morphology.
    
    Next, the predicted growth morphology and size throughout the transverse section of melt pool are demonstrated along with the experimental measurements for three sets of process parameters as high LED ($P$: 122 W,  $\vec{v}$: 50 mm/s, LED: 2.44 J/mm), medium LED ($P$: 96 W,  $\vec{v}$: 67 mm/s, LED: 1.43 J/mm), and low LED ($P$: 162 W,  $\vec{v}$: 957 mm/s, LED: 0.169 J/mm). Fig. \ref{fig:low_led_valid} demonstrates the predicted and measured microstructure at varying locations along the transverse section of the melt pool for $P$: 162 W,  $\vec{v}$: 957 mm/s. The calculated LED is 0.169 J/mm at this condition, which we identify as the low LED case. A conduction mode melt pool with the low depth-to-width aspect ratio is shown in the optical micrograph. The SEM micrographs in Fig. \ref{fig:low_led_valid} present a planar structure. From the thermal model, a very low $R$ with $R < Vcs$ is predicted near the edge of the melt pool, hence a planar structure is expected to form. in the low LED case, the values of $R$ vary within the range of $0.2-0.9 m/s$ (figure~\ref{fig:transverse_melt_pool_GR}), which falls into the region of $R > V_{ab}$. Note that, in experiments solidification always begins with a planar structure at the melt pool boundary ($R < V_{cs}$), which is followed by a transition to cellular ($V_{cs} < R < V_{ab}$) or planar structure ($R > V_{ab}$), depending on the laser processing conditions.

    Fig. \ref{fig:mid_led_valid} demonstrates the predicted and measured microstructure at varying locations along the transverse section of the melt pool for $P$: 96 W,  $\vec{v}$: 67 mm/s. The calculated LED is 1.43 J/mm at this condition, which we call as the medium LED case. A conduction to keyhole transition mode melt pool with the medium depth-to-width aspect ratio is shown in the optical micrograph. The SEM micrographs in Fig. \ref{fig:mid_led_valid} present the microstructure and size at three different locations: bottom (a), top-middle (b), and top-edge (c). All locations show cellular structure with measured PDAS of $0.39\pm0.1 \mu m$, $0.38\pm0.1 \mu m$, and $0.44\pm0.2 \mu m$, and the PF-predicted PDAS of $0.38 \mu m$, $0.32 \mu m$, and $0.42 \mu m$, respectively, indicating very good agreement between the simulation and experimental measurements. The predicted morphologies and PDAS at varying locations, from top to bottom and from top-middle to top-edge, are also presented in Fig. \ref{fig:mid_led_valid}. The results show that the PDAS decreases from the top to the bottom of the melt pool with the maximum and the minimum being $0.46 \mu m$ and $0.38 \mu m$, respectively. In contrast, the PDAS increases from the top-middle to the top-edge region with minimum and maximum being $0.32 \mu m$ and $0.42 \mu m$, respectively. The predictions show that the cooling rate decreases from top-middle to top-edge explaining the increase in PDAS.

    Fig. \ref{fig:high_led_valid} demonstrates the predicted and measured microstructure at varying locations along the transverse section of the melt pool for $P$: 122 W,  $\vec{v}$: 50 mm/s. The calculated LED is 2.44 J/mm at this condition, which we refer to as the high LED case. A keyhole mode melt pool with high depth-to-width aspect ratio is shown in the optical micrography image. The bottom region is exposed to a high $G$ and low $R < Vcs$, and hence corresponds to the planar region as shown in Fig. \ref{fig:GV_map}. On the other hand, the top region shows a relatively lower $G$ and $R$ (between $Vcs$ and $Vab$). Therefore, a cellular structure is stable in this region, as depicted in Fig. \ref{fig:GV_map}. The SEM micrographs in Fig. \ref{fig:high_led_valid} present the microstructure and size at three different locations: the bottom (a), top-middle (b), and top-edge (c). The bottom (a) location shows a planar structure. Similarly, a planar structure is predicted by the PF simulations at the corresponding location. The top-middle location (b) shows a cellular structure with the measured PDAS of $0.43\pm0.2 \mu m$. The PF-predicted PDAS at this location is $0.36 \mu m$. The top-edge (c) shows a long cellular structure which grow from the melt pool boundary to the interior melt pool. The measured PDAS at this location is $0.58\pm0.2 \mu m$ while the PF-predicted PDAS is $0.46 \mu m$. Both the PF-predicted and experimentally measured PDAS increase from the top-middle to the top-edge region similar to the medium LED case as shown in Fig. \ref{fig:mid_led_valid}.

    Above, we explained the variation in the morphology and size of the microstructural features across the solidified melt pool for different processing conditions. Nb microsegregation is another important phenomenon to discuss. The overall results indicate that as the cooling rate increases, the concentration of Nb in the intercellular region decreases. For example, a reduction in the cooling rate from $7.5\times10^5K/s$ to $1.5\times10^5K/s$ leads to an increase in the predicted amount of Nb in the intercellular region from 4 to 5 (in at.\% Nb). Also, once the growth rate reaches beyond the absolute stability limit, a segregation-free planar structure is formed. Since the growth rate is a strong function of the laser speed, it is assumed to have this type of segregation free structure to be formed under sufficiently high laser speed conditions.

    \subsection{Effect of Permeability} \label{effect_of_perm}
    
    As stated in the previous paragraphs, the interface permeability has a strong influence on the predictions of finite interface dissipation PF model. With a proper selection of interface permeability, it is possible to model either the systems with local equilibrium conditions or highly non-equilibrium conditions. A number of test simulations (1D and 2D) have been run for varying permeability values. 
    
     \begin{figure}[!ht]
     \centering
     \includegraphics[width=0.9\columnwidth]{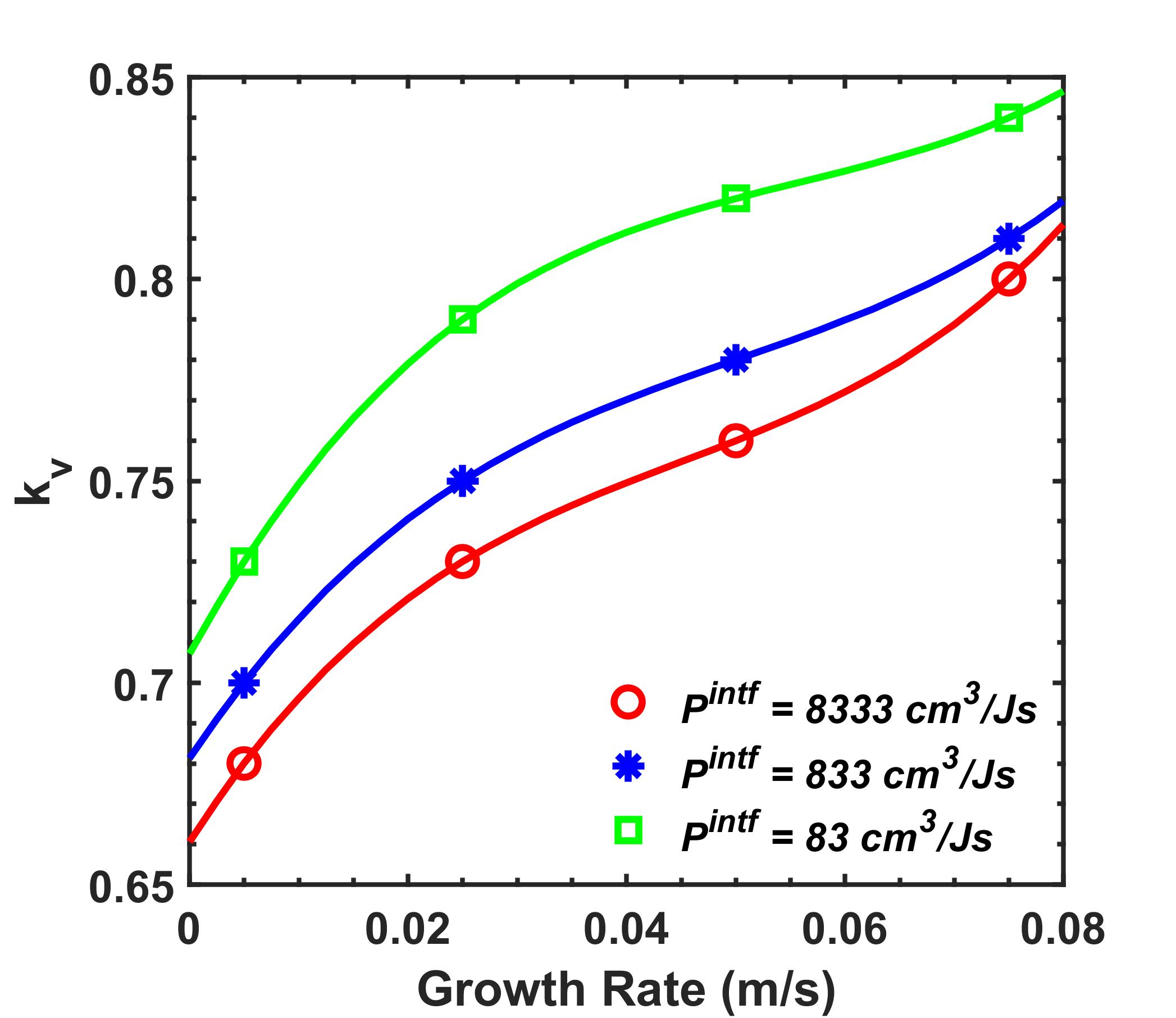}
     \caption{Comparison between the simulated segregation coefficient $k_{v}$ as a function of the growth rate $R$ in the binary Ni-Nb alloy with different interface permeability $P^{intf}$} 
     \label{fig:effect_of_perm}
    \end{figure}  

    The results indicated that the value of permeability might affect multiple features of microstructure including growth structure, PDAS, cellular/dendritic shape, microsegregation, and segregation coefficient.
    
    As an example, in this section, we investigate the effect of permeability on the segregation coefficient $k_v$. Figure \ref{fig:effect_of_perm} represents the variation in $k_v$ as a function of $R$ for varying permeability values as predicted from the 1D simulations of binary Ni-Nb alloy. Based on the material parameters listed in Table \ref{tab:mat_pars}, the calculated $P^{intf}$ value according to the equation given in Section \ref{PFM_sec} is on the order of $10^4$ $cm^3/Js$. 
    
    Steinbach et al. \cite{steinbach2012phase} stated that the permeability equation gives the upper limit, above which the model recovers to a conventional model with equal chemical potentials. In contrast, if lower values of permeability are utilized, the effect of solute trapping is enhanced indicated with higher values of $k_v$, which is the result of slower chemical partitioning process due to the low permeability across the solid/liquid interface. This is clearly shown in Fig. \ref{fig:effect_of_perm}. As the permeability decreases the predicted $k_v$ increases. In order to determine the actual value of permeability, the original paper \cite{steinbach2012phase} suggests comparing the calculated $k(v)$ with that measured experimentally. In the present work, since the experimental measurement of $k(v)$ is not available, the value of actual permeability have been determined by comparing other metrics such as PDAS or microsegregation.

    Here, it is worth mentioning that $k_v$ indeed depends on a combination of interface width and permeability. However, in the present work, the interface width was fixed to that was reported in Ghosh et al. \cite{ghosh2017primary} to be able to make reasonable comparisons of the models. Also note that, although the effect of interface width on $k_v$ was not the scope of the present work, from the test simulations with varying interface widths it was confirmed that the employed value ($\eta=$ $32 nm$) to give the best computational performance as well as the best agreement with the experimentally measured PDAS for the selected permeability value as given in Table \ref{tab:mat_pars}.

    \subsection{Comparison with Conventional PF Model}

    As stated already, conventional PF models predicting microstructural features of Ni-Nb alloy under AM conditions are available in the current literature.
    
        \begin{figure}[!ht]
     \centering
     \includegraphics[width=0.9\columnwidth]{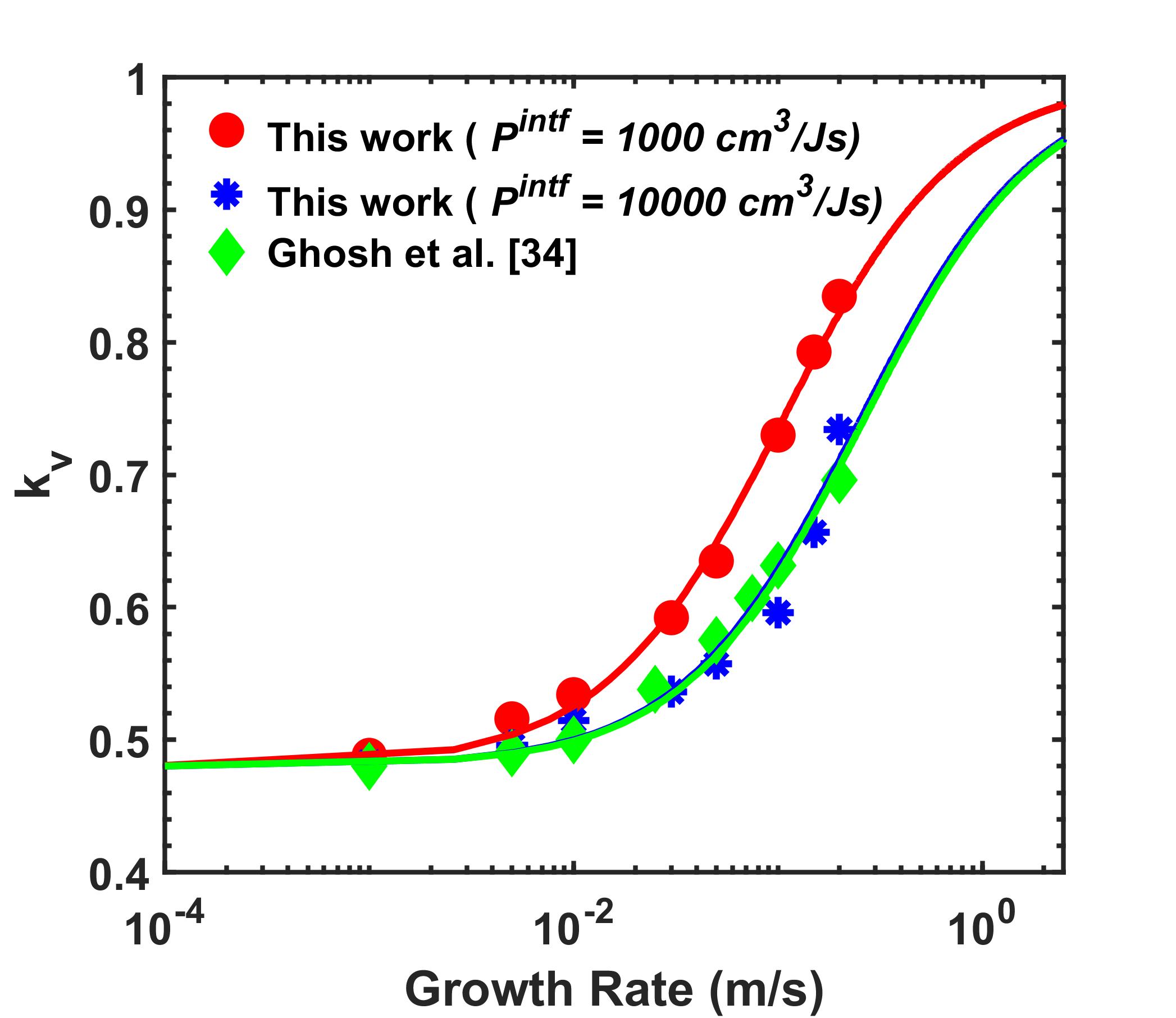}
     \caption{ Comparison between the simulated segregation coefficients $k_{v}$ from Ghosh et. al \cite{ghosh2017primary} (conventional PF model) and this work (finite interface dissipation PF model) in the quasi-binary Ni-Nb alloy} 
     \label{fig:model_comparisons}
    \end{figure}

    In the present work, the parameters employed in Ghosh et al. \cite{ghosh2017primary} have been adopted as much as possible to be able to compare the results from the present work with those from Ghosh et al. \cite{ghosh2017primary}.

    In this regard, first the predicted PDAS from each model will be compared. In Fig.~\ref{fig:high_led}, the predicted cooling rate varied from $2\times10^4 K/s$ to $3.5\times10^6 K/s$ as we go from the bottom to the top of the melt pool, and the predicted PDAS ranged from $0.7 \mu m$ to $0.3 \mu m$, respectively.

    A similar work by Ghosh et al. \cite{ghosh2017primary} predicted the cooling rate to range from $5\times10^4 K/s$ to $3\times10^6 K/s$, and corresponding PDAS to be within the range of $1.6\mu m$-$0.14\mu m$ under similar laser processing parameters ($P=195W$ and $\vec{v}=800mm/s$), using the conventional PF model under the local equilibrium assumption. It is clear that in the work by Ghosh et al. a larger range of PDAS were predicted. This difference can be attributed to multiple factors. First of all, although both studies use the binary Ni-3.2 at.\% Nb as the modeling material, the aforementioned work aimed to approximate the microstructure of Inconel 718, hence adopted the thermodynamic data from a quasi-binary phase diagram of Inconel 718. Based on the utilized phase diagram, the freezing range and equilibrium segregation coefficient were reported as 57 K \cite{nie2014numerical} and 0.48 \cite{nie2014numerical}, respectively in the aforementioned study, whereas these properties in the present work are given as 14 K \cite{saunders1998calphad} and 0.68 \cite{saunders1998calphad}, respectively. It is inevitable that these variations in the material properties will also influence the microstructure predictions.

    Therefore, this kind of comparison is not helpful to show how the finite interface dissipation model differs from the conventional model. To be able to directly compare the results, each model should adopt the same parameters. Although an apples-to-apples comparison is still not possible due to the model parameters specific to each model (e.g. ``interface mobility'' is only defined in the finite interface dissipation PF mode), by employing the same thermodynamic data and using the same values of common model parameters (e.g diffusivity, anisotropy coefficient, interface width) sufficiently reasonable comparisons can be made. 
    
    In this regard, 1-dimensional simulations have been run employing the same thermodynamic data (based on the quasi-binary Ni-Nb phase diagram) that is used in \cite{ghosh2017primary}, as well as utilizing the same model parameters as much as possible (as listed in Table \ref{tab:mat_pars}).
    
    The finite interface dissipation model enables simulating solidification both under local equilibrium conditions and non-equilibrium conditions, depending on the selected value of permeability. To demonstrate this, two values of permeability (i.e. low vs high) have been determined and the predicted values of $k_{v}$ have been compared. The results demonstrated that when a high value of permeability ($P^{intf}=10,000$ $cm^3/Js$) is used, the finite interface dissipation model recovers the conventional model results indicated with low values of $k_v$.
    
    In contrast, when a low value of permeability ($P^{intf}=1,000$ $cm^3/Js$) is used, higher kv values associated with non-equilibrium conditions can be modeled. Prdicting higher $k_v$ at the lower $P^{intf}$ values can be explained due to the fact that a lower $P^{intf}$ value can slow down the chemical partitioning process, hence enhances the effect of solute trapping during rapid solidification.

    \section{Summary and Conclusion}\label{sec:conc}
    In the present work, we developed a computational modeling framework by weakly coupling a finite element based thermal model to a non-equilibrium PF model to investigate the rapid solidification microstructure that forms during the single track melting of Ni-3.2 at.\% Nb. Most of the previous literature adopted PF models with the assumption of local equilibrium at the solid-liquid interface. However, this assumption is not valid for rapid solidification processes occuring under L-PBF conditions. In contrast, the present study employs the finite interface dissipation PF model with the ability of describing highly non-equilibrium systems, and is thus well suited to predict the rapid solidification microstructure during the L-PBF process.
    
    The present study provides a consistent framework by adopting a binary Ni-3.2 at.\% Nb alloy in both PF model and validation experiments. Note that the finite interface dissipation PF model adopted in the present work is well suited for describing multi-phase and multi-component alloy systems. The main goal of the present work was to elucidate the influence of the process parameters on the variability in solidification microstructure under single-track laser melting conditions. Once the microstructure variability in a simple binary alloy during single-track experiments is well understood, the next step would be employing the present framework to investigate more complex alloy systems (e.g. Inconel 718) under multi-track laser melting conditions. 
    
    The major findings of the present work can be summarized as follows:

    1. The single track laser melting simulations were run at varying laser power ($P: 70-255 [W]$) and laser speed ($\vec{v}: 50-2300 [mm/s]$). A planar to cellular transition was predicted as we go from the melt pool boundary to the interior of majority of the keyhole mode melt pools. A planar interface beyond the limit of absolute stability was observed to be predominant in the conduction mode melt pools with high growth rates. 
    
    2. A solidification map of the Ni-Nb alloy under L-PBF conditions (Fig.~\ref{fig:GV_map}) describing the variation in the morpholgy and size of cellular segregation structure as a function of temperature gradient $G$ and growth rate $R$ was presented. The minimum $G$ and $R$ conditions, leading to a segregation-free planar structure (beyond the absolute stability) was shown. Note that a solidification map describing the variation in the morphology (cellular, planar, and mixed) as a function of laser power $P$ and scan speed $\vec{v}$ will be presented in a following work.
    
    3. The typical cooling rate during L-PBF ranged from $~10^5 K/s$ to $~10^8 K/s$ depending on the process parameters and the location in the melt pool. Above $3.5\times10^6 K/s$, a transition from cellular to planar interface was predicted with growth rates beyond $Vab$.
    
    4. The predicted morphology and size of microstructure during the single-track melt pool solidification under varying process conditions were compared against the experimental measurements, and very good agreement was achieved regarding the cell spacing (ranging between ~0.2 and ~0.5 $\mu$m) and  the amount of Nb in the intercellular region (from 4-5 at.\% Nb). 
    
    5. The effect of permeability on the predicted segregation coefficient ($k_v$) as a function of the growth rate was demonstrated. The results indicated that with the lower values of permeability, higher values of $k_v$ were predicted, indicating enhanced solute trapping effect.
    
    6. The finite interface dissipation PF model results were compared with those predicted from a conventional PF model \cite{ghosh2017primary}. It was shown that with a low value of permeability, the former model recovers to the latter with local equilibrium condition (indicated with low values of $k_v$). On the other hand, employing a higher value of permeability in the former model resulted in a higher $k_v$, associated with non-equilibrium conditions.

    Finally, we note that the authors have recently developed a 3-dimensional PF model (see Fig. \ref{fig:3d_sim}) and compared the PDAS predicted from 2-dimensional model (2D) with those predicted from 3-dimensional (3D) model. A slight difference (e.g. $0.4 \mu m$ vs $0.46 \mu m$) in the predicted PDAS was shown. A thorough analysis in which a comparison of microstructural features under varying processing conditions predicted using 2D and 3D models will be presented in future work. Furthermore, the incorporation of the Lattice Boltzmann method to the developed PF model is currently being carried out by the authors in order to simulate the growth of cellular segregation structure in the presence of melt convection, which was shown to be an important factor in some literature work \cite{rojas2015phase,takaki2015gpu} .

\section*{Acknowledgements}
The authors acknowledge the support of NASA through grant NNX15AD71G. The experimental aspect of this work was carried out with support from the Army Research Office (ARO) under Contract No. W911NF-18-1-0278. LJ wishes to acknowledge the support of the National Science Foundation through award NSF-DGE 1545403 (NRT-DESE: Data-Enabled Discovery and Design of Energy Materials). KK and RA also acknowledge the Lawrence Livermore National Laboratory under Collaborative R\&D in Support of LLNL Missions, Task Order No. B623252 and Master Task Agmt. B575363. VA and RA also acknowledge support from NSF through grant No. CMMI-1462255. The FE and PF simulations were carried out at the Texas A\&M Super Computing Center.

\FloatBarrier

\appendix
\label{appendix-sec1}
\section{}
\setcounter{figure}{0}

    \begin{figure*}[!ht]
      \centering
      \includegraphics[width=0.9\textwidth]{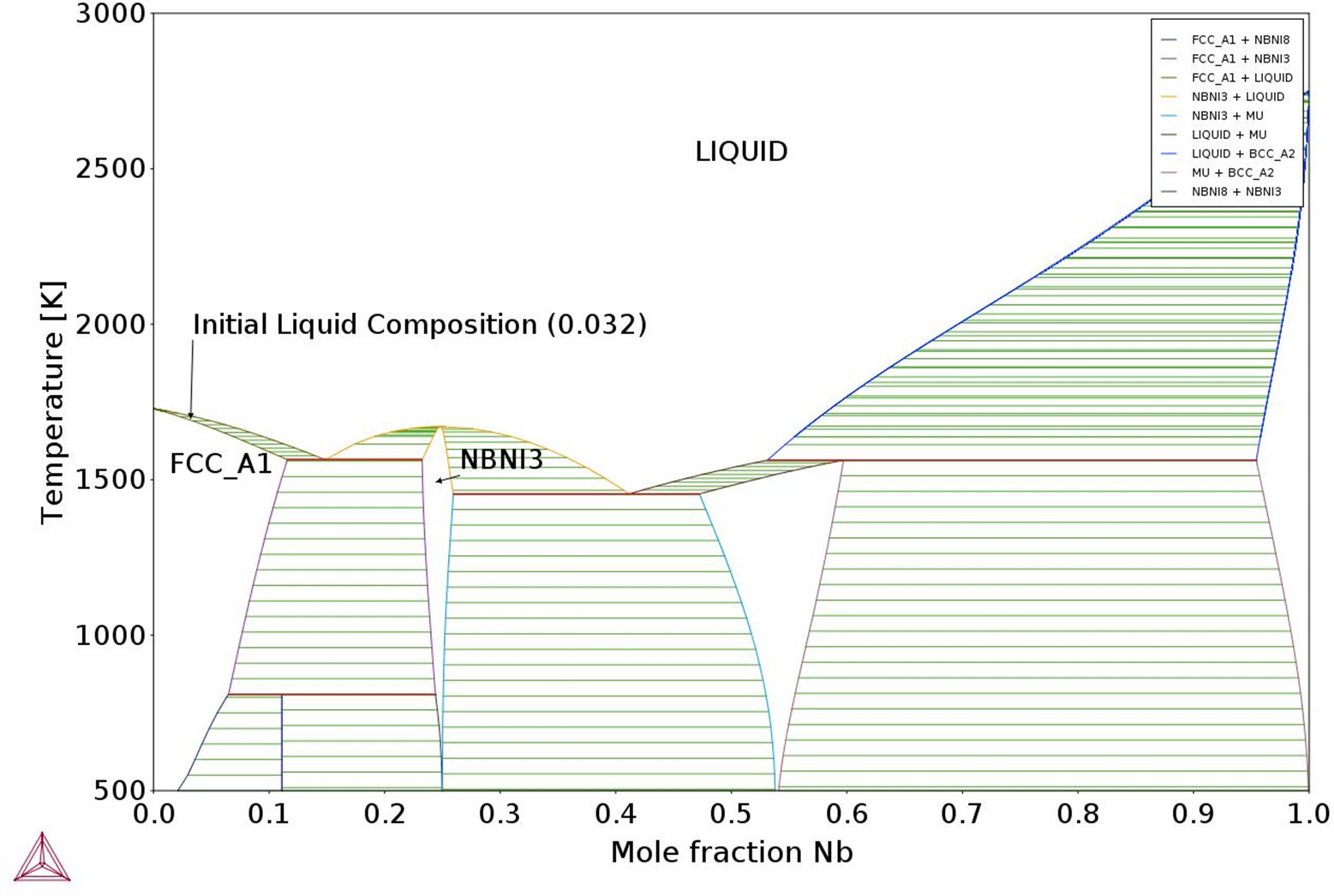}
      \caption{The phase diagram of binary Ni-Nb alloy based on the thermodynamics data given in \cite{joubert2004assessment}}
      \label{fig:phase_diagram}
    \end{figure*}     
    
    \begin{figure*}[!ht]
      \centering
      %\subfloat[]{\includegraphics[width=0.45\textwidth]{images/after_revision/3D_blue_cropped.jpeg}}
      %\subfloat[]{\includegraphics[width=0.5\textwidth]{images/after_revision/2D_blue_v2_cropped.jpeg}}
      \includegraphics[width=0.8\textwidth]{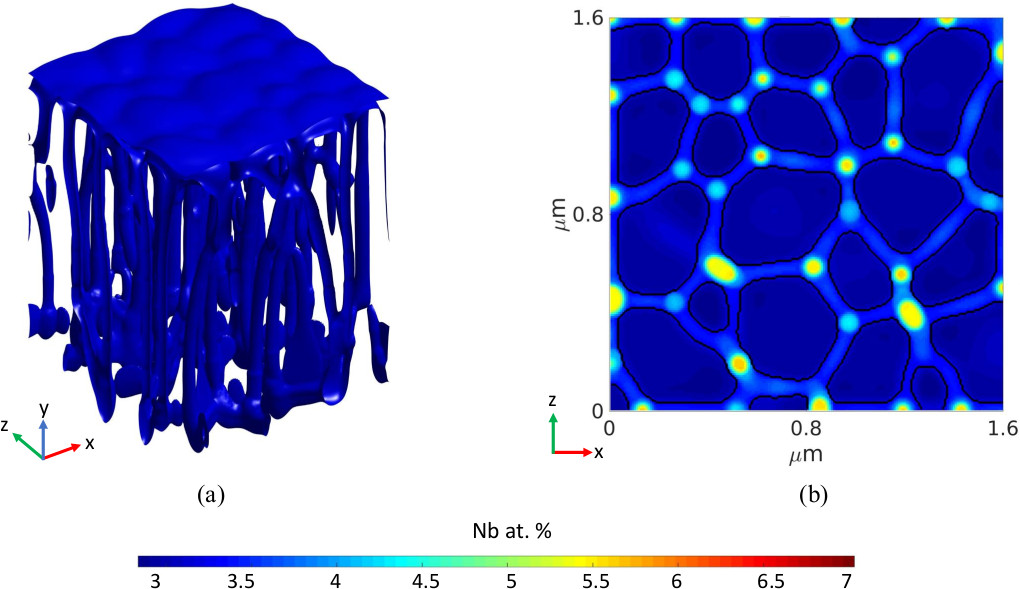}      
      %\subfloat[]{\includegraphics[width=0.5\textwidth]{images/new2/2D_blue_v2.jpeg}} \hspace*{8pt}  \begin{minipage}{\linewidth}\vskip-150pt \rotatebox{90}{mole fraction Nb} \end{minipage}
      %\put(10,10){hello}
      \caption{Two different views of growing cells (i.e. oblique (a) and transverse section(b)) obtained from 3-dimensional PF model is presented. The colorbar indicates the concentration of Nb in atomic percentage. }
      \label{fig:3d_sim}
    \end{figure*}

%% References


\bibliographystyle{elsarticle-num}

\bibliography{sample}

\end{document}